\documentclass[a4paper,11pt]{article}
\pdfoutput=1 

\usepackage{jcappub} 

\usepackage[T1]{fontenc} 
\usepackage{dsfont}

\newcommand{\id}{\mathds{1}}
\newcommand{\inte}{\!\int\! \text{d}}

\newcommand{\Scal}{\mathcal{S}}

\newcommand{\Pcal}{\mathcal{P}}
\newcommand{\Ncal}{\mathcal{N}}
\newcommand{\T}{\scalebox{0.6}{\text{T}}}
\newcommand{\smallindex}[1]{{\scriptscriptstyle #1}}
\newcommand{\I}{\smallindex{I}}
\newcommand{\J}{\smallindex{J}}
\newcommand{\K}{\smallindex{K}}
\renewcommand{\L}{\smallindex{L}}
\newcommand{\A}{\smallindex{A}}
\newcommand{\B}{\smallindex{B}}
\newcommand{\C}{\smallindex{C}}
\newcommand{\D}{\smallindex{D}}
\NewDocumentCommand{\e}{mmm}{%
  e^{\scriptscriptstyle #1}_{{\scriptscriptstyle #2}\,#3}%
}

\title{\boldmath Multi-gravity cosmology beyond pairwise interactions}

\author[a,b,1]{J. Flinckman,\note{Corresponding author.}}

\affiliation[a]{Stockholm University,\\Roslagstullsbacken 21,\\SE-106 91 Stockholm, Sweden}
\affiliation[b]{The Oskar Klein Centre,\\AlbaNova University Centre,\\SE-106 91 Stockholm, Sweden }

\emailAdd{Joakim.flinckman@fysik.su.se}

\abstract{
We develop the cosmological framework for Hassan--Schmidt-May multi-gravity, a ghost-free extension of bimetric theory with genuinely non-pairwise interactions of multiple spin-2 fields. For metrics that are simultaneously homogeneous and isotropic, the background dynamics reduce to a modified Friedmann equation supplemented by a system of coupled quartic polynomial equations for the scale-factor ratios. The spin-2 interaction contributes to the Friedmann equation as an additional effective fluid. We identify a regular high-density branch that approaches the standard radiation- or matter-dominated Universe, while regular low-density branches approach proportional vacuum solutions, where the interaction acts as a cosmological constant and the leading correction to the effective spin-2 energy density provides an additional matter contribution. We also determine the relative null-cone structure: on the regular high-density branch, the additional null cones approach wider opening angles than the matter-coupled null cone, while all null cones coincide in the vacuum limit.
At linear order, we solve the algebraic equations determining the non-metric vielbein perturbations, derive the first-order consistency equations and their scalar and vector projections, and obtain the complete system of coupled tensor perturbation equations. These results establish a framework for studying cosmological solutions and perturbations in ghost-free multi-gravity beyond bimetric theory and its pairwise-interacting extensions.
}

\begin{document}
\maketitle
\flushbottom

\section{Introduction}
\label{sec:intro}

While the standard cosmological model, derived from Einstein's theory of general relativity, has been remarkably successful, some of the biggest open questions in modern physics are tied to the theory of gravity. In particular, the dark matter problem \cite{Rubin:1980zd,Clowe:2006eq,Bertone:2016nfn}, the accelerated expansion of the Universe \cite{SupernovaSearchTeam:1998fmf,SupernovaCosmologyProject:1998vns} and the Hubble tension \cite{Planck:2018vyg, Riess:2021jrx} might each, by themselves, justify modifying our theory of gravity. What remains less clear is which modifications are warranted.

In particle physics, multiplet structures are ubiquitous and constitute an essential part of the Standard Model where, for example, the electroweak bosons transform under internal gauge symmetries. Indeed, all fundamental interactions except gravity share this feature, naturally raising the question of whether a modified theory of gravity might admit an analogous structure: Could general relativity likewise be embedded in a broader framework, just as classical electromagnetism was ultimately understood as one component of electroweak theory? This would necessitate more than one gravitational field or graviton, leading to the idea of multi-gravity. The spin-2 case is, however, far less explored than its lower-spin counterparts, and already at the classical level, such theories generically lead to Boulware–Deser-like ghost instabilities \cite{Boulware:1972yco,Boulware:1972zf,Creminelli:2005qk}, greatly restricting the admissible interactions.

Following the construction of a ghost-free theory of massive gravity \cite{deRham:2010ik,deRham:2010kj,Hassan:2011hr,Hassan:2011tf,Hassan:2011ea,Hassan:2012qv,Comelli:2013txa}, Hassan and Rosen constructed the first ghost-free theory of two interacting gravitational metrics, commonly known as bi-gravity or bimetric theory \cite{Hassan:2011zd,Hassan:2011ea,Hassan:2017ugh,Hassan:2018mbl}. This is a theory of two non-linearly interacting spin-2 fields, one massless and one massive \cite{Hassan:2012wr}. While some extensions of bi-gravity can trivially be constructed from pairwise interactions \cite{Khosravi:2011zi,Molaee:2019knc,Dokhani:2020jxb,Wood:2025ujj}, it was not until the construction proposed by Hassan and Schmidt-May \cite{Hassan:2018mcw} that a genuine multi-gravity theory with the potential to be ghost-free was formulated. They showed that their interaction avoids a known obstruction in multi-gravity and admits constraints essential for consistency. The existence of the additional ghost-eliminating constraints was demonstrated in \cite{Flinckman:2025bje}, thereby establishing that the theory is free of Boulware–Deser ghosts. It thus provides a nonlinear theory of one massless and multiple massive spin-2 fields \cite{Flinckman:2024zpb}. Even in this broader setting, the spin-2 fields are not related by an internal gauge symmetry, so neither bimetric nor multi-gravity theory realises a multiplet structure analogous to those found in particle physics, although both give rise to a rich phenomenology.

Bimetric cosmologies are well studied; see, for example \cite{vonStrauss:2011mq,Akrami:2012vf,Koennig:2013fdo,Comelli:2011zm,Volkov:2011an,Akrami:2015qga,Hogas:2021lns,Luben:2018ekw,Hogas:2021saw}. They admit viable cosmologies with features such as dynamical dark energy and may even alleviate the Hubble tension \cite{Hogas:2025ahb}. The additional massive spin-2 field is also a viable cold dark matter candidate \cite{Aoki:2016zgp, Babichev:2016hir, Babichev:2016bxi}. However, in bi-gravity the additional spin-2 field cannot naturally provide both dark energy and dark matter. If its mass is large, it is a viable dark matter candidate, but the background is essentially indistinguishable from $\Lambda$CDM. If instead the mass is of the order of the Hubble scale, the interaction drives a dynamical dark energy component, but the field is possibly too light to account for dark matter. With multiple massive spin-2 fields of multi-gravity, there is a wide mass spectrum \cite{Flinckman:2024zpb}, so that a heavy sector may contribute to the dark matter budget while a light sector simultaneously accounts for the accelerated expansion. This possibility motivates a systematic investigation of the cosmological dynamics of multi-gravity.

In this work, we initiate the study of cosmologies in multi-gravity beyond pairwise interactions by developing the background field equations and the perturbative constraints required for a systematic analysis of cosmological perturbations. Section \ref{sec:MG-theory} introduces the ghost-free multi-gravity theory, its covariant field equations, and the associated Lorentz and Bianchi constraints. In Section \ref{sec:FRLW}, we impose a homogeneous and isotropic ansatz, solve the background constraints, and reduce the cosmological dynamics to a modified Friedmann equation supplemented by a system of coupled quartic equations for the scale-factor ratios. We also analyse the high- and low-density limits and the corresponding null-cone configurations. Section \ref{sec:pert_eq} develops the linear vielbein perturbations, solves the linear Lorentz constraints, and derives a convenient form of the perturbative Bianchi constraints. Finally, we derive the tensor perturbation equations and the scalar and vector projections of the constraints, while leaving a complete analysis of the dynamical scalar and vector sectors to future work.

\section{Multi-gravity theory}
\label{sec:MG-theory}

The multi-gravity action is formulated in terms of $\Ncal$ spacetime metrics $g^{\I}_{\mu \nu}$, 
each with a corresponding Einstein–Hilbert term, coupled through a non-derivative interaction potential,
\begin{align}
\label{action}
    \mathcal{S} = \inte^4 x\!\left[\,\sum_{I=1}^\Ncal \tfrac{1}{2}m_{\I}^2 \sqrt{-g_{\I}^{}}\left( R_{\I} - 2 \Lambda_{\I} \right) - V(e_1,\dots, e_{\Ncal})\right] +\sum_{I=1}^{\Ncal} \Scal_{\text{M}}^{\I}[e_{\I}, \psi_{\I}].
\end{align}
$m_{\I}$ and $\Lambda_{\I}$ are Planck-like masses and cosmological constants associated with each metric. The potential $V$ is an antisymmetric product of the vielbeins $\e{A}{I}{\mu}$ associated with the metrics,
\begin{align}
\label{HR_Pot}
    V = \frac{m^4}{4!}\!\!\sum_{\I\J\K\L=1}^{\Ncal} \!\!\beta^{\smallindex{IJKL}}\delta_{\!\smallindex{ABCD}}^{\alpha \beta \gamma \delta}\,\e{A}{I}{\alpha}\e{B}{J}{\beta}\e{C}{K}{\gamma}\e{D}{L}{\delta}, && g^\I_{\mu \nu}= e^\A_{\I \, \mu}\eta^{\,}_{\A\B}e^\B_{\I\, \nu},
\end{align}
where $\delta_{\!\smallindex{ABCD}}^{\alpha \beta \gamma \delta} = \epsilon_{\smallindex{ABCD}}\epsilon^{\alpha \beta \gamma \delta}$ is the generalised delta, $\epsilon_{0123}=\epsilon^{0123}=+1$ and $\beta^{\I \J \K \L}$ is a dimensionless, totally symmetric interaction parameter. Each metric can interact with its own matter sector $\Scal_{\text{M}}^{\I}$, but a matter field cannot interact with more than one metric or directly with fields in other sectors. The Standard Model matter fields therefore couple to only one metric, whose geometry determines the geodesics followed by Standard Model particles.

Each Einstein–Hilbert term is covariant under diffeomorphisms and local Lorentz transformations of the vielbein, but the interaction breaks this down to the diagonal subgroup in which each vielbein and its associated metric transform under the same transformation,
\begin{align}
    \e{A}{I}{\mu}(x) \mapsto L^{\A}_{~\B}(x)\e{B}{I}{\nu}(x)\frac{\partial x^\nu}{\partial x'^\mu}, && g^{\I}_{\mu \nu}(x) \mapsto \frac{\partial x^\alpha}{\partial x'^\mu}g^{\I}_{\alpha \beta}(x)\frac{\partial x^\beta}{\partial x'^\nu},
\end{align}
where $L^{\T}\eta L = \eta$.

The potential \eqref{HR_Pot} was first proposed in \cite{Hinterbichler:2012cn} and for $\Ncal =2$, it provides the vielbein form of the ghost-free bimetric interaction. For $\Ncal>2$, the theory is, however, not ghost-free for generic $\beta^{\I\J\K\L}$ \cite{Afshar:2014dta, deRham:2015cha}. In fact, the only ghost-free multi-field interaction beyond pairwise bimetric interactions is obtained when $\beta^{\I \J \K \L}$ is rank one \cite{Flinckman:2026non},
\begin{align}
    \beta^{\I \J \K \L} = \beta^{\I}\beta^{\J}\beta^{\K}\beta^{\L},
\end{align}
which results in the Hassan–Schmidt-May potential \cite{Hassan:2018mcw} and can be written in the form,
\begin{align}
\label{HSM_pot}
    V =\frac{m^4}{4!}\delta^{\alpha \beta \gamma \delta}_{\!\smallindex{ABCD}}\,u^{\A}_{~\alpha}u^{\B}_{~\beta}u^{\C}_{~\gamma}u^{\D}_{~\delta} = m^4 \det u, && u^{\A}_{~\mu}= \sum_{I=1}^{\Ncal}\beta^{\I}\e{A}{I}{\mu},
\end{align}
where $u$ is invertible and we will use the notation $u^\alpha_{~\A}=(u^{-1})^\alpha_{~\A}$. We will also assume that all $\beta^\I$ have the same sign and are non-vanishing.\footnote{If $\beta^\I=0$, the corresponding vielbein decouples from the interaction. Moreover, the couplings may be absorbed into the rescaling $e_\I\rightarrow\beta^\I e_\I$, together with appropriate rescalings of $m_\I$ and $\Lambda_\I$. A negative $\beta^\I$ then reverses the timelike leg, $e^0_{\I\,\mu}\rightarrow-e^0_{\I\,\mu}$, as part of the transformation $e^A_{\I\,\mu}\rightarrow-e^A_{\I\,\mu}$. This transformation is proper but not orthochronous, so frames with couplings of opposite signs have opposite time orientations.}

The Hassan–Schmidt-May theory has been argued to be ghost-free \cite{Hassan:2018mcw, Flinckman:2025bje} and to propagate one massless spin-2 field and $\Ncal{-}1$ massive spin-2 fields at the quadratic level \cite{Flinckman:2024zpb}. The field equations, $\frac{\delta \mathcal{S}}{\delta e^\mu_{\I\A}}\eta^{}_{\A\B}e^{\B}_{\I \, \nu} =0$, take the form of a modified set of Einstein equations,
\begin{align}
\label{eom}
    G^{\I}_{\mu \nu} + \Lambda^{\;}_{\I} g^{\I}_{\mu\nu}  = \frac{1}{m_\I^2}\Big[ T^{\I}_{\mu \nu}+V^{\I}_{\mu \nu}\Big], \qquad I = 1,\ldots, \Ncal,
\end{align}
where $G^{\I}_{\mu \nu}=R^{\I}_{\mu \nu}-\tfrac{1}{2}R^{\I}g^\I_{\mu \nu}$ is constructed from the Levi–Civita connection ${}^\I \nabla$ associated with $g^\I_{\mu\nu}$, satisfying ${}^{\I}\nabla_\mu g^\I_{\alpha\beta}=0$. The matter and interaction contributions are,
\begin{align}
    T^{\I}_{\mu\nu}&=-\frac{1}{\sqrt{-g_{\I}}}\frac{\delta \Scal_{\mathrm M}^{\I}}{\delta e^\mu_{\I\,\smallindex{A}}}\eta_{\smallindex{AB}}\e{B}{I}{\nu},\\    
\label{eom_pot}
    V^{\I}_{\mu \nu}&=\frac{1}{\sqrt{-g_{\I}}}\frac{\delta V}{\delta e^\mu_{\I\, \smallindex{A}}}\eta_{\smallindex{AB}}\e{B}{I}{\nu} =- m^4\beta^{\I}\det(e_{\I}^{-1}u)\e{A}{I}{\mu}u^\sigma_{~\A}g^{\I}_{\sigma \nu},
\end{align}
where the contribution $V^{\I}_{\mu \nu}$ can be interpreted as the energy-momentum of non-minimally coupled spin-2 fields.

Since the metric, the Einstein tensor and energy-momentum are symmetric, the field equations \eqref{eom} imply,
\begin{align}
\label{symcond}
    V_{[\mu \nu]}^{\I} &=0 &&\Longrightarrow  &&
    \e{A}{I}{[\mu}\eta^{\,}_{\A\B}u^{\B}_{\; \nu]} = \sum_{J=1}^{\Ncal}\beta^{\J}\e{A}{I}{[\mu}\eta_{\A\B}\e{B}{J}{\nu]}=0.
\end{align}
These are $6\Ncal$ algebraic equations and correspond to the field equations for the non-dynamical Lorentz degrees of freedom contained in the vielbeins \cite{Hinterbichler:2012cn, Flinckman:2025bje}. They are therefore called the Lorentz constraints. These are not all independent, as is evident from multiplying by $\beta^{\I}$ and summing over $I$, yielding the identity,
\begin{align}
    \sum_{I=1}^\Ncal \beta^{\I}\e{A}{I}{[\mu}\eta^{\,}_{\A\B}u^{\B}_{\; \nu]} =u^{\A}_{\; [\mu}\eta^{\,}_{\A\B}u^{\B}_{\; \nu]}=0.
\end{align}
The Lorentz constraints \eqref{symcond} thus provide $6(\Ncal-1)$ algebraic equations, enough to determine the $6(\Ncal-1)$ relative Lorentz fields of the vielbeins so that only the metric variables remain. In contrast to bimetric theory and pairwise-interacting multimetric theories, there is, however, no known metric formulation of the Hassan–Schmidt-May interaction.

There is yet another set of constraints that can be obtained directly from the covariant field equations \eqref{eom}. Taking the $I$th covariant derivative of the $I$th field equation and using ${}^{\I}\nabla_\mu G^\mu_{\I\, \nu}={}^{\I}\nabla_\mu T^\mu_{\I\, \nu}=0$ yields,
\begin{align}
\label{bianchi}
    {}^{\I}\nabla_{\mu} V^\mu_{\I \, \nu}=0.
\end{align}
These relations are first order in derivatives and are thus constraints, which we will call Bianchi constraints. These are not all independent, since diffeomorphism invariance implies the identity,
\begin{align}
    \sum_\I \sqrt{-g_\I}\,{}^\I\nabla_\mu V^\mu_{\I\,\nu}=0,
\end{align}
given that the Lorentz constraints $V^\I_{[\mu \nu]}=0$ have been imposed.

The Lorentz and Bianchi constraints can be used to eliminate non-dynamical fields. Initially, there are $16\Ncal$ independent components $\e{A}{I}{\mu}$, and \eqref{symcond} and \eqref{bianchi} eliminate $6(\Ncal{-}1)$ and $4(\Ncal{-}1)$ of these, leaving,
\begin{align}
    16\Ncal - 6(\Ncal-1) -4(\Ncal-1) = 6(\Ncal-1) + 16.
\end{align}
Using local Lorentz invariance and general covariance, we fix an overall local Lorentz frame and the coordinates, eliminating another $6+4$ fields. General covariance also gives four constraints, corresponding to the Hamiltonian and momentum constraints in the canonical formulation. This generically leaves us with,
\begin{align}
    5(\Ncal-1) + 2 + (\Ncal-1),
\end{align}
dynamical fields, where the first two terms count the polarisations of one massless spin-2 field and $\Ncal{-}1$ massive spin-2 fields, while the remaining term corresponds to $\Ncal{-}1$ potential Boulware–Deser-like ghosts. Due to the particular structure of the Hassan–Schmidt-May interaction, a canonical analysis has shown that there are $\Ncal-1$ additional constraints which eliminate the ghosts \cite{Flinckman:2025bje}, leaving this a ghost-free theory of interacting spin-2 fields.

\section{Background cosmology}
\label{sec:FRLW}

\subsection{Homogeneous and isotropic ansatz}

We now consider homogeneous and isotropic solutions of the field equations \eqref{eom}. A priori, the metrics need not be simultaneously homogeneous and isotropic in the same coordinate frame, but we restrict our attention to an ansatz for which they are, namely,
\begin{align}
\label{e=Le}
    \e{A}{I}{\mu}&= L^{\A}_{\I \,\B}\overline{e}^{\B}_{\I \, \mu},\\
\label{cosmo_ansatz}
    \overline{e}{}^{\B}_{\I \, \mu} &= \text{diag}\Big(N_{\I}(t),\, \tfrac{a_{\I}(t)}{\sqrt{1-kr^2}},\,a_{\I}(t) r,\, a_{\I}(t) r\sin\theta\Big),
\end{align}
where $L^{\scalebox{0.65}{\textit{A}}}_{\I \,\scalebox{0.65}{\textit{B}}}$ is a local Lorentz transformation, $N_{\I}(t)>0$ are lapse functions and $a_{\I}(t)>0$ are scale factors for each of the vielbeins. This yields the standard FLRW line elements,
\begin{align}
\label{metric_ansatz}
    \text{d}s_{\I}^2 &= -N^2_{\I}(t)\text{d}t^2 + a^2_{\I}(t)\gamma_{ij}\text{d}x^i\text{d}x^j, \\
\label{spatial_background}
    \gamma_{ij}\text{d}x^i\text{d}x^j &= \tfrac{\text{d}r^2}{1-kr^2} + r^2(\text{d}\theta^2 + \sin^2\theta\text{d}\phi^2),
\end{align}
where we have chosen the spatial curvatures $k$ to be the same for all line elements.\footnote{This is not an additional restriction. If one initially allows distinct curvatures $k_{\I}$, spatial isotropy of the field equations requires $V^r_{\I\,r}=V^\theta_{\I\,\theta}$, which implies $(k_{\I}-k_{\J})\sum_{\K}\beta^{\K}a_{\K}=0$. Since invertibility of $u$ requires $\sum_{\K}\beta^{\K}a_{\K}\neq0$, all interacting sectors must have the same spatial curvature, $k_{\I}=k_{\J}$.}

Substituting the ansatz \eqref{e=Le} into the Lorentz constraints \eqref{symcond} gives,
\begin{align}
    \sum_\J \beta^{\J} \overline{e}^{\C}_{\I\, [\mu}L^{\A}_{\I\, \C}\eta_{\A\B}L^{\B}_{\J\, \D}\overline{e}^{\D}_{\J\, \nu]}=0.
\end{align}
It can be easily verified that setting all $L^\A_{\I\, \B}$ equal to common Lorentz transformation $L^\A_{\I\, \B}=L^\A_{~\B}$ gives a solution, since each,
\begin{align}
    \overline{e}^{\C}_{\I\, \mu}L^{\A}_{~\C}\eta_{\A\B} L^{\B}_{~\D}\overline{e}^{\D}_{\J\, \nu}=\overline{e}^{\A}_{\I\, \mu}\eta_{\A\B} \overline{e}^{\B}_{\J\, \nu},
\end{align}
is trivially symmetric. Since the theory is invariant under an overall local Lorentz transformation, the common Lorentz factor $L^{\A}_{~\B}$ drops out of the action and does not enter the field equations. 

With the Lorentz fields eliminated, we can derive the cosmological field equations. The Einstein tensors take the form,
\begin{align}
\label{G00}
    G^0_{\I \, 0} &= -\frac{3}{a_{\I}^2}\bigg[\frac{\dot{a}_{\I}^2}{N_{\I}^2}+ k \bigg],\\
    G^i_{\I \, j} &=- \bigg[\frac{1}{N_{\I}^2}\bigg\lbrace\! \bigg(\frac{\dot{a}_{\I}}{a_{\I}} \bigg)^2 +\frac{2}{a_{\I}}\bigg( \ddot{a}_{\I}-\frac{\dot{N}_{\I}}{N_{\I}}\dot{a}_{\I}\bigg)\! \bigg\rbrace + \frac{k}{a_{\I}^2}\bigg]\delta^i_j.
\end{align}
It will be convenient to introduce the notation,
\begin{align}
    S_a = \sum_\I \beta^\I a_\I, && S_N = \sum_\I \beta^\I N_\I,
\end{align}
so that the potential takes the form,
\begin{align}
\label{pot_SS3}
    V=m^4\det u=m^4\sqrt{\gamma}S_NS_a^3.
\end{align}
For non-degenerate $u$ we must therefore have $S_a\neq 0$ and $S_N\neq0$. By a straightforward computation, the interaction \eqref{eom_pot} reads,
\begin{align}
\label{V00}
    V^0_{\I \, 0} &=- b^{\I} \bigg[\sum_\J \beta^{\J} \frac{a_{\J}}{a_{\I}} \bigg]^3 = -b^\I \left(\frac{S_a}{a_\I }\right)^3,\\
\label{Vij}
    V^i_{\I \, j} &= -b^{\I}\sum_\K\beta^{\K} \frac{N_{\K}}{N_{\I}}\bigg[\sum_\J \beta^{\J} \frac{a_{\J}}{a_{\I}} \bigg]^2\delta^i_j =- b^\I \left(\frac{S_a}{a_\I}\right)^2 \frac{S_N}{N_\I} \delta^i_j,
\end{align}
where $b^{\I} = \beta^{\I} m^4$. Since the spatial parts $G^i_{\I \, j}$ and $V^i_{\I\, j}$ are proportional to the identity, it is convenient to write $G^i_{\I \, j}= G^S_\I \delta^i_j$ and $V^i_{\I \, j}=V^S_\I\delta^i_j$, where,
\begin{align}
\label{GS}
    G_{\I}^S= - \bigg[\frac{1}{N_{\I}^2}\bigg\lbrace\! \bigg(\frac{\dot{a}_{\I}}{a_{\I}} \bigg)^2 +\frac{2}{a_{\I}}\bigg( \ddot{a}_{\I}-\frac{\dot{N}_{\I}}{N_{\I}}\dot{a}_{\I}\bigg)\! \bigg\rbrace + \frac{k}{a_{\I}^2}\bigg], &&
    V_{\I}^S = - b^\I \left(\frac{S_a}{a_\I}\right)^2 \frac{S_N}{N_\I}.
\end{align}

\subsection{Background Bianchi constraints and cosmological branches}

Before writing down the field equations, we first evaluate the Bianchi constraints \eqref{bianchi}. Their spatial components vanish identically,
\begin{align}
    {}^{\I}\nabla_\mu V^\mu_{\I \; i}&= \partial_\mu V^\mu_{\I \;i} + {}^{\I}\Gamma^{\mu}_{\; \mu \nu}V^\nu_{\I \; i}- {}^{\I}\Gamma^\nu_{\; \mu i}V^{\mu}_{\I \;\nu}\notag\\
    &=\partial_i V^S_{\I} + {}^{\I}\Gamma^0_{\; 0 i}\big[V^S_{\I} -V^0_{\I\, 0}\big]=0,
\end{align}
since $V^0_{\I\,i}=V^i_{\I\,0}=0$, and the final equality follows from homogeneity, $\partial_i V^S_{\I}=0$, together with the standard identity ${}^{\I}\Gamma^0_{~0i}=0$ for the FLRW metric.

By a similar computation, the temporal component yields,
\begin{align}
\label{temp_bianchi}
    {}^{\I}\nabla_\mu V^\mu_{\I\,0}=\partial_0V^0_{\I\,0}+3\frac{\dot a_{\I}}{a_{\I}}\left[V^0_{\I\,0}-V^S_{\I}\right]=0,
\end{align}
which is non-trivial and provides constraints on, for example, the lapse functions $N_\I$.

If we interpret the interaction contribution to each $V^\mu_{\I \,\nu}$ as an effective fluid, 
\begin{align}
\label{spin-2_w}
    \widetilde{\rho}_{\I}&=-V^0_{\I\,0}=b^{\I}\left(\frac{S_a}{a_{\I}}\right)^3, &
    \widetilde{p}_{\I}&=V^S_{\I}=-b^{\I}\left(\frac{S_a}{a_{\I}}\right)^2\frac{S_N}{N_{\I}}, &
    \widetilde{w}_{\I}&=\frac{\widetilde{p}_{\I}}{\widetilde{\rho}_{\I}}=-\frac{a_{\I}S_N}{N_{\I}S_a},
\end{align}
the temporal Bianchi constraint \eqref{temp_bianchi} takes the form of a continuity equation,
\begin{align}
\label{spin2_continuity}
    \dot{\widetilde{\rho}}_{\I}+3\frac{\dot a_{\I}}{a_{\I}}\left(\widetilde{\rho}_{\I}+\widetilde{p}_{\I}\right)=0.
\end{align}
Evaluating $\dot{\widetilde{\rho}}_{\I}$ gives,
\begin{align}
    \dot{\widetilde{\rho}}_{\I}=3\widetilde{\rho}_{\I}\left[\frac{\dot S_a}{S_a}-\frac{\dot a_{\I}}{a_{\I}}\right],
\end{align}
and \eqref{spin2_continuity} reduces to,
\begin{align}
\label{cosmo_bianchi}
    S_a^2\left[\dot S_a-\frac{\dot a_{\I}}{N_{\I}}S_N\right]=0.
\end{align}
This equation has two branches of solutions. The first, which we call the algebraic branch, is,
\begin{align}
\label{alg_bianchi}
    S_a=\sum_{\J}\beta^{\J}a_{\J}=0,
\end{align}
which clearly necessitates either $\beta^\I<0$ or $a_\I<0$ for some $I$, thereby violating the assumptions $a_\I>0$ and that all $\beta^\I$ have the same sign. However, on this branch, $\widetilde{\rho}_{\I}=\widetilde{p}_{\I}=0$, the potential \eqref{pot_SS3} vanishes and the homogeneous and isotropic theory reduces to $\Ncal$ copies of standard FLRW cosmology. We will later see that perturbations around this branch contain strongly coupled modes.\footnote{This branch lies outside the non-degenerate sector in which $u^{-1}$ appearing in \eqref{eom_pot} exist. It may nevertheless be obtained by expressing the interaction equations through $\text{adj}(u) = \det(u)u^{-1}$, which vanishes on the present FLRW branch.}

For non-degenerate $u$, $V\neq0$ implies $S_a\neq0$, and \eqref{cosmo_bianchi} instead gives,
\begin{align}
    \frac{\dot a_{\I}}{N_{\I}}=\frac{\dot S_a}{S_N}=\frac{\sum_{\J}\beta^{\J}\dot a_{\J}}{\sum_{\K}\beta^{\K}N_{\K}}.
\end{align}
Since the right-hand side is independent of $I$, it follows that,
\begin{align}
    \frac{\dot a_1}{N_1}=\frac{\dot a_2}{N_2}=\dots=\frac{\dot a_{\Ncal}}{N_{\Ncal}}.
\end{align}
This determines all but one of the lapses $N_{\I}$ in terms of a residual lapse, which we take without loss of generality to be $N_1$. We call this the dynamical branch, with solutions,\footnote{This parametrisation assumes $\dot a_1\neq0$. This is not a restriction on the solutions: the Bianchi constraints imply $\dot a_{\I}/N_{\I}=\dot a_1/N_1$ for all $I$, so if $\dot a_1$ vanishes at an isolated turning point, all $\dot a_{\I}$ vanish simultaneously. The relative lapses may then be defined by continuity across that point.}

\begin{align}
\label{dyn_bianchi}
    N_{\I}=N_1\frac{\dot a_{\I}}{\dot a_1}.
\end{align}

\subsection{Friedmann equations}
With the Lorentz and Bianchi constraints solved and all but one of the lapses determined, we can now for simplicity use the time-reparametrisation symmetry to work in a time coordinate where $N_1=1$, i.e. $\text{d}t' = N_1(t)\text{d}t$. For convenience, we drop the prime $'$ and introduce the scale-factor ratios $y_{\I}=a_{\I}/a_1$, with $y_1=1$. We also suppress the sector label on quantities associated with the first sector $I=1$, writing $\widetilde\rho_1= \widetilde\rho$, $\widetilde{p}_1 = \widetilde p$, $\widetilde w_1 = \widetilde w$ etc. Introducing the Hubble parameter $H= \dot{a}/a$ and defining,
\begin{align}
\label{c_def}
    c_{\I}&=\frac{N_{\I}}{y_{\I}}=1+\frac{\mathrm{d} \ln y_{\I}}{\mathrm{d} \ln a}, && c_1=1,
\end{align}
we can simplify (\ref{G00}–\ref{Vij}) on the dynamical branch, so that,
\begin{align}
    G^0_{\I\,0}&=-\frac{3}{y_{\I}^2}\left[H^2+\frac{k}{a^2}\right], &
    G^S_{\I}&=-\frac{1}{y_{\I}^2}\left[H^2+\frac{k}{a^2}+\frac{2}{c_{\I}}\left(\dot H+H^2\right)\right],\\
    \widetilde{\rho}_{\I}&=\frac{b^{\I}}{y_{\I}^3}S_y^3, &
    \widetilde{p}_{\I}&=\widetilde{w}_{\I}\widetilde{\rho}_{\I}=\frac{b^{\I}}{y_{\I}^3}\frac{\widetilde{w}}{c_{\I}}S_y^3,
\end{align}
where, in analogy to $S_a$ and $S_N$, we have defined $S_y$ so that,
\begin{align}
\label{SNtoSy}
    S_y = \sum_\I \beta^\I y_\I, && S_N&=\sum_{\J}\beta^{\J}y_{\J}c_{\J} = -\widetilde{w}S_y.
\end{align}
This relates the spin-2 fluids so that,
\begin{align}
    \widetilde{\rho}_{\I}=\frac{b^{\I}}{b^1y_{\I}^3}\widetilde{\rho}, &&
    \widetilde{p}_{\I}=\frac{b^{\I}}{b^1y_{\I}^3c_{\I}}\widetilde{p}= \frac{b^{\I}\widetilde{w}}{b^1y_{\I}^3c_{\I}}\widetilde{\rho}, && \widetilde{w}_I = \widetilde{w}/c_I.
\end{align}
We now couple the first sector, $I=1$, to a perfect fluid with equation of state $p=w\rho$. Its energy-momentum tensor reads $T^\mu{}_\nu=\rho[(1+w)v^\mu v_\nu+w\delta^\mu{}_\nu]$, where $v^\mu$ is the fluid 4-velocity. The 00-component of the field equations for $I=1$ then yields the modified Friedmann equation,
\begin{align}
\label{friedmann_1}
    H^2+\frac{k}{a^2}-\frac{\Lambda_1}{3}=\frac{1}{3m_1^2}\big[\rho+\widetilde{\rho}\big], && \widetilde{\rho}=b^1S_y^3.
\end{align}
 The acceleration equation for $I=1$ takes the familiar form,
\begin{align}
\label{acc_1}
    \dot{H}=\frac{k}{a^2}-\frac{1}{2m_1^2}\Big[ (1+ w)\rho+(1+\widetilde{w})\widetilde{\rho} \Big],
\end{align}
where we have used \eqref{friedmann_1} to eliminate $H^2$. Thus, for the metric coupled to matter, the interaction contribution enters the Friedmann and acceleration equations as an additional fluid with energy density $\widetilde{\rho}$ and a time-dependent equation-of-state parameter $\widetilde{w}$.

If no matter is coupled to the remaining sectors, their field equations take the form,
\begin{gather}
\label{friedmann_I}
    H^2+\frac{k}{a^2}-\frac{\Lambda_{\I}y_{\I}^2}{3}=\frac{\alpha_{\I}}{3m_1^2y_{\I}}\widetilde{\rho}, \quad\qquad \alpha_{\I}=\frac{m_1^2b^{\I}}{m_{\I}^2b^1}\\
\label{acc_I}
    \dot{H}=\frac{k}{a^2}+\frac{\Lambda_{\I}y_{\I}^2}{3}\left[c_{\I}-1\right]-\frac{\alpha_{\I}\widetilde{\rho}}{3m_1^2y_{\I}}\left[1+\frac{c_{\I}}{2}\left(1+3\widetilde{w}_{\I}\right)\right].
\end{gather}
We can now obtain an algebraic equation by taking the difference between \eqref{friedmann_1} and \eqref{friedmann_I},
\begin{align}
\label{1I_diff}
    \Lambda_1-\Lambda_{\I}y_{\I}^2+\frac{1}{m_1^2}\left[\rho+\widetilde{\rho}-\frac{\alpha_{\I}}{y_{\I}}\widetilde{\rho}\right]=0.
\end{align}
This is an equation for the scale-factor ratios $y_{\I}$ in terms of the theory parameters $\alpha_{\I}$, $\Lambda_{\I}$ and the matter energy density $\rho$.

The full background dynamics can therefore be characterised by the modified Friedmann equation \eqref{friedmann_1}, the matter continuity equation $\dot{\rho}+3H(\rho + p)=0$ together with the $\Ncal-1$ quartic polynomial equations obtained by multiplying \eqref{1I_diff} by $y_\I$,
\begin{align}
\label{friedmann}
    H^2+\frac{k}{a^2}-\frac{\Lambda_1}{3}&=\frac{1}{3m_1^2}\left[\rho+\widetilde{\rho}(y_{\J})\right],\\
\label{friedmann_poly}
    \mathcal{P}_{\I}(y_{\J},\rho)&=m_1^2\Lambda_{\I}y_{\I}^3-\left[m_1^2\Lambda_1+\rho+\widetilde{\rho}(y_{\J})\right]y_{\I}+\alpha_{\I}\widetilde{\rho}(y_{\J})=0.
\end{align}
Since $\Pcal_1$ is trivial, \eqref{friedmann_poly} forms a system of $\Ncal-1$ polynomial equations $\mathcal{P}_{\I}=0$ for $y_{\I}$.

Physical solutions for $y_\I$ should result in positive scale factors $a_\I>0$ for all $I$. In Appendix \ref{app:positive_cosmological_solutions}, we demonstrate that, when all $\beta^\I$ have the same sign, as also required by the common orientation of the vielbeins, the system admits positive solutions for $y_\I$, thereby guaranteeing $a_\I>0$.

\subsection{Asymptotic regimes}

The asymptotic cosmological behaviour of $\widetilde{\rho}$ can be understood directly from the algebraic equations \eqref{friedmann_poly}. Making the dependence of the interaction energy density on the scale-factor ratios explicit, they read,
\begin{align}
\label{friedmann_poly_explicit}
    \mathcal{P}_{\I}(y_{\J},\rho)=m_1^2\Lambda_{\I}y_{\I}^3-\Big[m_1^2\Lambda_1+\rho+b^1\Big(\sum_{\J}\beta^{\J}y_{\J}\Big)^3\Big]y_{\I}+\alpha_{\I}b^1\Big(\sum_{\J}\beta^{\J}y_{\J}\Big)^3=0.
\end{align}
Consider first the high-density limit $\rho\rightarrow\infty$. If the scale-factor ratios remain bounded, the term $\rho y_{\I}$ dominates unless $y_{\I}\rightarrow0$. In this limit, $\sum \beta^\I y_\I\rightarrow\beta^1$, so the leading terms in \eqref{friedmann_poly_explicit} give,
\begin{align}
    -\rho y_{\I}+\alpha_{\I}m^4(\beta^1)^4\simeq0, && y_{\I}\simeq\frac{\alpha_{\I}m^4(\beta^1)^4}{\rho}=\frac{m_1^2m^4\beta^{\I}(\beta^1)^3}{m_{\I}^2\rho}.
\end{align}
This defines the regular high-density branch. On this branch,
\begin{align}
    S_y=\beta^1+\mathcal{O}(\rho^{-1}), && \widetilde{\rho}=m^4(\beta^1)^4+\mathcal{O}(\rho^{-1}), && \widetilde{w}&=-1+\mathcal{O}(\rho^{-1}).
\end{align}
The spin-2 energy density therefore remains finite while $\rho$ diverges, and hence $\widetilde{\rho}/\rho\rightarrow0$. The early Universe consequently approaches the standard radiation- or matter-dominated cosmology, with a subdominant cosmological-constant-like spin-2 contribution.

A second high-density branch arises when one or more scale-factor ratios become large. If $y_{\I}\sim\rho^l$ and no cancellation suppresses $S_y$, then,
\begin{align}
    \rho y_{\I}&\sim\rho^{1+l}, & \widetilde{\rho}(y_{\J})y_{\I}&\sim\rho^{4l}, & y_{\I}^3\sim\widetilde{\rho}(y_{\J})&\sim\rho^{3l}.
\end{align}
Balancing the first two terms requires $l=1/3$. The leading part of \eqref{friedmann_poly_explicit} then gives,
\begin{align}
    \rho+\widetilde{\rho}(y_{\J})\simeq0, \qquad \widetilde{\rho}\simeq-\rho, \qquad \widetilde{w}\rightarrow w.
\end{align}
This is a screening branch on which the spin-2 contribution tracks and cancels the dominant matter density in both the Friedmann and acceleration equations. The expansion is therefore controlled by subleading contributions and does not approach the standard high-density radiation- or matter-dominated cosmology. Note, however, that $\widetilde{\rho}\simeq-\rho$ is only possible if one relaxes the assumption that all $\beta^\I$ have the same sign.

In the low-density limit $\rho\rightarrow0$, the polynomial equations reduce to,
\begin{align}
\label{vacuum_polynomial}
    \mathcal{P}_{\I}(y_{\J}, \rho=0)=m_1^2\Lambda_{\I}y_{\I}^3-\left[m_1^2\Lambda_1+\widetilde{\rho}(y_{\J})\right]y_{\I}+\alpha_{\I}\widetilde{\rho}(y_{\J})=0.
\end{align}
These are time-independent quartic equations whose solutions have constant scale-factor ratios $y_{\I}=\bar y_{\I}$. For each such solution, $\overline{S}_y=\sum_{\J}\beta^{\J}\bar y_{\J}$ is constant, and hence,
\begin{align}
    \widetilde{\rho}=\rho_{\mathrm{CC}}=b^1\overline{S}_y^3, &&\widetilde{w}=-1.
\end{align}
The spin-2 interaction therefore acts exactly as a cosmological constant, with,
\begin{align}
    \Lambda_{\mathrm{eff}}=\Lambda_1+\frac{\rho_{\mathrm{CC}}}{m_1^2}.
\end{align}
Depending on the sign of $\rho_{\mathrm{CC}}$, the interaction can either enhance or reduce the bare cosmological constant $\Lambda_1$. But given the assumption that $\beta^\I$ all have the same sign, $\rho_{\mathrm{CC}}$ is necessarily positive.

We next consider a small but non-vanishing matter density. Since $\rho$ enters \eqref{friedmann_poly_explicit} as a small change in the coefficients of the algebraic equations, a regular vacuum root shifts smoothly away from its vacuum value. The leading correction can therefore be written as,
\begin{align}
    y_{\I}=\bar y_{\I}+\zeta_{\I}\rho+\mathcal{O}(\rho^2),
\end{align}
where the coefficients $\zeta_{\I}$ are determined by the theory parameters and the chosen vacuum solution. It follows that,
\begin{align}
    S_y=\overline{S}_y+\rho\sum_{\I\neq1}\beta^{\I}\zeta_{\I}+\mathcal{O}(\rho^2),
\end{align}
and consequently,
\begin{align}
    \widetilde{\rho}(y_{\J})&=\rho_{\mathrm{CC}}+C\rho+\mathcal{O}(\rho^2), & C&=3b^1\overline{S}_y^2\sum_{\I\neq1}\beta^{\I}\zeta_{\I}.
\end{align}
Using the effective continuity equation together with $\rho\propto a^{-3(1+w)}$, the corresponding pressure is,
\begin{align}
    \widetilde{p}=-\rho_{\mathrm{CC}}+Cw\rho+\mathcal{O}(\rho^2).
\end{align}
The leading spin-2 contribution therefore behaves as dark energy, while the first correction tracks the equation of state of the dominant matter component. For example, during matter domination, $w=0$ and $\rho=\rho_{\mathrm m}\propto a^{-3}$, so that,
\begin{align}
    \widetilde{\rho}&=\rho_{\mathrm{CC}}+C\rho_{\mathrm m}+\mathcal{O}(\rho_{\mathrm m}^2), & \widetilde{p}&=-\rho_{\mathrm{CC}}+\mathcal{O}(\rho_{\mathrm m}^2).
\end{align}
For $\rho_{\mathrm{CC}}>0$ and $C>0$, the spin-2 sector therefore behaves at the background level as a combination of positive dark energy and an additional pressureless dust contribution. The term $C\rho_{\mathrm m}$ can mimic a dark-matter contribution to the background expansion. It is not, however, an independent matter component at first order, as its density remains algebraically tied to $\rho_{\mathrm m}$. The late-time Friedmann equation consequently becomes,
\begin{align}
    H^2+\frac{k}{a^2}=\frac{\Lambda_{\mathrm{eff}}}{3}+\frac{1+C}{3m_1^2}\rho+\mathcal{O}(\rho^2).
\end{align}
At the background level, this is degenerate with an additional pressureless component, or equivalently with a rescaling of the effective gravitational coupling to $\rho_\text{m}$.

The constant part of the spin-2 interaction generates the curvature scale,
\begin{align}
    \Lambda_{\mathrm{spin}}=\frac{\rho_{\mathrm{CC}}}{m_1^2}=\frac{\beta^1m^4\overline{S}_y^3}{m_1^2}.
\end{align}
Thus, even when the bare cosmological constant vanishes, $\Lambda_1=0$, a positive $\rho_{\mathrm{CC}}$ can drive late-time accelerated expansion.

\subsection{Null-cone structure}

Before considering perturbations of the homogeneous and isotropic solutions of (\ref{friedmann}–\ref{friedmann_poly}), we briefly comment on the null-cone structure of the theory. On the dynamical branch, after fixing $N_1=1$, the remaining lapses are,
\begin{align}
    N_{\I}
    =y_{\I}c_{\I}.
\end{align}
The corresponding metrics can therefore be written as,
\begin{align}
    \mathrm{d}s_{\I}^2
    =y_{\I}^2\left[-c_{\I}^2\mathrm{d}t^2
    +a^2\gamma_{ij}\mathrm{d}x^i\mathrm{d}x^j\right].
\end{align}
The null-cones are thus all coaxial but with different opening angles. Introducing conformal time $\mathrm{d}\eta=\mathrm{d}t/a$ and the radial coordinate $\mathrm{d}\chi=\mathrm{d}r/\sqrt{1-kr^2}$, radial null curves satisfy,
\begin{align}
    \frac{\mathrm{d}\chi}{\mathrm{d}\eta}=\pm c_{\I}.
\end{align}
Thus $c_{\I}$ \eqref{c_def} is the propagation velocity associated with the $I$th null cone relative to the matter-coupled $I=1$ sector, for which $c_1=1$. 

On the high-density branch where $y_{\I}\to0$ and $S_y\to\beta^1$, and for a dominant fluid $\rho\propto a^{-3(1+w)}$ it follows that,
\begin{align}
    c_{\I}\longrightarrow4+3w.
\end{align}
The relative propagation velocity therefore approaches $c_{\I}=5$ during radiation domination and $c_{\I}=4$ during matter domination. The null cones of the additional metrics are therefore wider than the ``physical null cone'' associated with the metric that defines our geometry and couples to the Standard Model matter sector, whose energy-momentum tensor is $T^\mu_{~\nu}$.

In the low-density limit, $y_{\I}\rightarrow\bar y_{\I}$ is constant, so that $c_{\I}\rightarrow1$ and all metrics become conformally related,
\begin{align}
    \mathrm{d}s_{\I}^2=\bar y_{\I}^{2}\left[-\mathrm{d}t^2+a^2\gamma_{ij}\mathrm{d}x^i\mathrm{d}x^j\right],
\end{align}
and their null cones coincide. For the nearby solution $y_{\I}=\bar y_{\I}+\zeta_{\I}\rho+\mathcal{O}(\rho^2)$, using $\rho\propto a^{-3(1+w)}$ gives,
\begin{align}
    c_{\I}  =1-3(1+w)\frac{\zeta_{\I}}{\bar y_{\I}}\rho+\mathcal{O}(\rho^2).
\end{align}
The null cones thus differ from the common vacuum cone only by corrections linear in the matter density, which vanish as $\rho\rightarrow0$.

\section{Cosmological perturbations}
\label{sec:pert_eq}
We now consider perturbations around the cosmological backgrounds described above.

\subsection{Perturbative vielbein expansion}
The Einstein tensors can be perturbed using standard methods in terms of the metric perturbations $\delta g^{\I}_{\mu\nu}=g^{\I}_{\mu\nu}-\overline{g}^{\I}_{\mu\nu}$. However, the interaction potential must be perturbed at the level of the vielbeins, including their relative Lorentz degrees of freedom. A convenient formalism for this was developed in \cite{Flinckman:2024zpb}, in which the vielbein perturbations are parametrised by the metric perturbations together with additional Lorentz fields $\omega^\I_{\mu \nu}=-\omega^\I_{\nu\mu}$. To first order, the vielbein perturbation can then be written,
\begin{align}
\label{delta_1}
    \delta e^{\A}_{\I \, \mu}=\tfrac{1}{2}\overline{e}^{\A}_{\I \, \nu}\Big[ \delta g^\nu_{\I\, \mu}-\omega^\nu_{\I\, \mu}\Big],
\end{align}
and the sum of vielbeins $u^{\A}_{\; \mu}=\sum_\I \beta^\I e^{\A}_{\I\, \mu}$ can similarly be written,
\begin{align}
\label{u_pert}
    u^{\A}_{\; \mu}=\overline{u}{}^{\A}_{\; \mu} + \delta u^{\A}_{\; \mu}+ \dots, && \delta u^{\A}_{\; \mu} = \sum_\I \beta^{\I} \delta e^{\A}_{\I \, \mu}.
\end{align}
We can now expand the potential contribution as,
\begin{align}
    V_{\mu}^{\I\,\nu}
    =\overline{V}{}_{\mu}^{\I\,\nu}
    +\delta V_{\mu}^{\I\,\nu}+\dots,
\end{align}
where, using \eqref{HSM_pot}, one finds,
\begin{align}
\label{deltaV}
    \delta V_{\mu}^{\I\,\nu}
    =-\tfrac{1}{2}\overline{V}{}_{\mu}^{\I\,\nu}\delta g^\sigma_{\I\,\sigma}
    +\delta e^{\A}_{\I\,\mu}\overline{e}^\sigma_{\I\,\A}\overline{V}{}_{\sigma}^{\I\,\nu}
    +2\overline{V}{}_{\mu\phantom{\A}}^{\I\,[\nu}\overline{u}\,{}^{\sigma]}_{\;\A}
    \delta u^\A_{\;\sigma}.
\end{align}
On the algebraic branch \eqref{alg_bianchi}, $\sum_\J\beta^{\J}a_{\J}=0$, both the background interaction contribution $\overline{V}_{\mu}^{\I\,\nu}$ and the linear perturbation $\delta V^{\I \, \nu}_\mu$ vanish. The linearised field equations therefore reduce to $\Ncal$ decoupled copies of the linearised Einstein equations, propagating only $2\Ncal$ modes. This is fewer than the $2+5(\Ncal-1)$ modes propagated by the full nonlinear theory \cite{Flinckman:2025bje} and recovered at the quadratic level around proportional backgrounds \cite{Flinckman:2024zpb}, leaving $3(\Ncal-1)$ modes absent from the linearised equations. The cubic term in the expansion of the interaction potential is, however, non-vanishing on this branch, so the interaction reappears at second order in the field equations. The missing $3(\Ncal-1)$ modes therefore have vanishing quadratic kinetic terms and enter only through nonlinear interactions, demonstrating that the algebraic branch is strongly coupled.\footnote{Analogous algebraic, also called self-accelerating, branches have previously been studied in bimetric and massive gravity \cite{vonStrauss:2011mq,Cusin:2015tmf,DeFelice:2012mx}. The bimetric branch develops a late-time tensor instability, while the massive-gravity branch contains a nonlinear ghost. These pathologies motivated the construction of so-called minimal theories of massive gravity and bigravity \cite{DeFelice:2015hla,DeFelice:2020ecp}.} Here we instead focus on the dynamical branch \eqref{dyn_bianchi}, which does not exhibit the same vanishing of the linear interaction.

\subsection{Perturbative Lorentz constraints}

Before we consider the perturbative field equations, we will focus on the linear Lorentz and Bianchi constraints, starting with the former. At linear order, the symmetrisation condition $V^{\I}_{[\mu \nu]}=0$ reduces to,
\begin{align}
    \delta V^{\I}_{[\mu \nu]}=0.
\end{align}
However, it is more convenient to linearise the equivalent form \eqref{symcond}, which reads,
\begin{align}
    u^{\A}_{\,[\mu}\eta_{\A\B}e^{\B}_{\I\,\nu]}=\overline{u}^{\A}_{\,[\mu}\eta_{\A\B}\overline{e}^{\B}_{\I\,\nu]}+\delta u^{\A}_{\,[\mu}\eta_{\A\B}\overline{e}^{\B}_{\I\,\nu]}+\overline{u}^{\A}_{\,[\mu}\eta_{\A\B}\delta e^{\B}_{\I\,\nu]}+\mathcal{O}(\delta e^2).
\end{align}
With the background symmetrisation conditions $\overline{u}^{\A}_{\,[\mu}\eta_{\A\B}\overline{e}^{\B}_{\I\,\nu]}=0$ imposed, we obtain, to first order,
\begin{align}
    \delta u^{\A}_{\,[\mu}\eta_{\A\B}\overline{e}^{\B}_{\I\,\nu]}+\overline{u}^{\A}_{\,[\mu}\eta_{\A\B}\delta e^{\B}_{\I\,\nu]}=0. \label{linear_lc}
\end{align}
Performing a $3+1$ split and inserting the homogeneous and isotropic background vielbeins $\overline{u}^{\A}_{~\mu}$ and $\overline{e}^{\A}_{\I\,\mu}$, the non-trivial components become,
\begin{align}
    (0i):\quad & \frac{S_a}{a_{\I}}\big[\delta g^{\I}_{i0}-\omega^{\I}_{i0}\big]-\sum_{\J}\beta^{\J}\frac{a_{\I}}{a_{\J}}\big[\delta g^{\J}_{i0}-\omega^{\J}_{i0}\big]-\frac{S_N}{N_{\I}}\big[\delta g^{\I}_{0i}-\omega^{\I}_{0i}\big]+\sum_{\J}\beta^{\J}\frac{N_{\I}}{N_{\J}}\big[\delta g^{\J}_{0i}-\omega^{\J}_{0i}\big]=0, \label{LC_vec}\\
    (ij):\quad & \sum_{\J}\beta^{\J}\frac{a_{\I}}{a_{\J}}\big[\delta g^{\J}_{[ij]}-\omega^{\J}_{[ij]}\big]+\frac{S_a}{a_{\I}}\big[\delta g^{\I}_{[ji]}-\omega^{\I}_{[ji]}\big]=0.
\end{align}
Since the metric perturbations $\delta g^{\I}_{\mu\nu}$ are symmetric, $\delta g^{\J}_{[ij]}=0$, the spatial equation admits the common solution,
\begin{align}
    \omega^i_{\I\,j}=\omega^i_{~j},
\end{align}
for all $I$, where the spatially lowered Lorentz fields are defined by,
\begin{align}
    \omega^{\I}_{ij}=\overline{g}^{\I}_{ik}\omega^k_{\I\,j}=a_{\I}^2\gamma_{ik}\omega^k_{\I\,j}.
\end{align}
Here and later it will be convenient to parametrise the metric perturbations as,
\begin{align}
    \delta g^\I_{00}&=-2N^2_\I\phi_\I, & \delta g^0_{\;0}&=2\phi_\I, & \delta g^{00}&=-\frac{2\phi_\I}{N^2_\I},\\
\label{delta_g0i}
    \delta g^\I_{i0}&=\delta g^\I_{0i}=N_\I a_\I n^\I_i, & \delta g^i_{\I\,0}&=\frac{N_\I}{a_\I}n^i_\I, & \delta g^0_{\I\,j}&=-\frac{a_\I}{N_\I}n^\I_j, & \delta g_\I^{i0}&=-\frac{1}{N_\I a_\I}n^i_\I,  \\
    \delta g^\I_{ij}&=a^2_\I h^\I_{ij}, & \delta g^i_{\;j}&=h^i_{\;j}, & \delta g^{ij}&=\frac{1}{a^2_\I}h^{ij},
\end{align}
where $n^i_\I=\gamma^{ij}n^\I_j$, $h^i_{\I\,j}=\gamma^{ik}h^\I_{kj}$, $h^{ij}_\I=\gamma^{ik}\gamma^{jl}h^\I_{kl}$ and $\gamma_{ij}$ is given by \eqref{spatial_background}. We similarly introduce the Lorentz boost vectors $A_i^{\I}$ through the non-vanishing spatiotemporal components of $\omega^{\I}_{\mu\nu}$,
\begin{align}
\label{omega0i}
    \omega^{\I}_{i0}&=-\omega^{\I}_{0i}=N_{\I}a_{\I}A_i^{\I}, & \omega^i_{\I\,0}&=\frac{N_{\I}}{a_{\I}}A^i_{\I}, & \omega^0_{\I\,j}&=\frac{a_{\I}}{N_{\I}}A_j^{\I}, 
\end{align}
where $A^i_{\I}=\gamma^{ij}A_j^{\I}$. If we now impose the background Bianchi constraints \eqref{dyn_bianchi} and use the $c_\I$ notation \eqref{c_def}, the spatiotemporal Lorentz constraint \eqref{LC_vec} reads,
\begin{align}
    \sum_{\J}\beta^{\J}y_{\J}\left(c_{\I}+c_{\J}\right)\left(A_i^{\I}-A_i^{\J}\right)=\sum_{\J}\beta^{\J}y_{\J}\left(c_{\I}-c_{\J}\right)\left(n_i^{\I}+n_i^{\J}\right).
\end{align}
These equations determine $\Ncal-1$ relative Lorentz boost vectors $A_i^{\I}-A_i^{\J}$. The remaining common boost is undetermined as a consequence of the overall local Lorentz invariance.

Using $\widetilde{w}_{\I}=\widetilde{w}/c_{\I}$ and \eqref{SNtoSy}, the expanded Lorentz constraint can be written as,
\begin{align}
    \left(1-\widetilde{w}_{\I}\right)A_i^{\I}=\left(1+\widetilde{w}_{\I}\right)n_i^{\I}+\frac{1}{S_y}\sum_{\J}\beta^{\J}y_{\J}\left(A_i^{\J}+n_i^{\J}\right)+\frac{\widetilde{w}_{\I}}{S_y}\sum_{\J}\frac{\beta^{\J}y_{\J}}{\widetilde{w}_{\J}}\left(A_i^{\J}-n_i^{\J}\right).
\end{align}
For generic backgrounds with $\widetilde{w}\neq0$, it is useful to parametrise the two combinations containing all boost fields by,
\begin{align}
    \lambda_i&=\frac{1}{2S_y}\sum_{\J}\beta^{\J}y_{\J}\left[A_i^{\J}+n_i^{\J}-\frac{1}{\widetilde{w}_{\J}}\left(A_i^{\J}-n_i^{\J}\right)\right],\\
    \mathcal{B}_i&=\frac{1}{2S_y}\sum_{\J}\beta^{\J}y_{\J}\left[A_i^{\J}+n_i^{\J}+\frac{1}{\widetilde{w}_{\J}}\left(A_i^{\J}-n_i^{\J}\right)\right].
\end{align}
Equivalently,
\begin{align}
    \sum_{\J}\beta^{\J}y_{\J}\left(A_i^{\J}+n_i^{\J}\right)&=S_y\left(\mathcal{B}_i+\lambda_i\right),\\
    \sum_{\J}\frac{\beta^{\J}y_{\J}}{\widetilde{w}_{\J}}\left(A_i^{\J}-n_i^{\J}\right)&=S_y\left(\mathcal{B}_i-\lambda_i\right).
\end{align}
The Lorentz constraint then takes the simple form,
\begin{align}
\label{boost_solution_general}
    A_i^{\I}=\lambda_i+\frac{1+\widetilde{w}_{\I}}{1-\widetilde{w}_{\I}}\left(n_i^{\I}+\mathcal{B}_i\right),
\end{align}
where $\mathcal{B}_i$ implicitly depends on the $A_i^\I$. Under a Lorentz boost $A_i^{\I}\rightarrow A_i^{\I}+\xi_i$, the two quantities transform as,
\begin{align}
    \lambda_i&\rightarrow\lambda_i+\xi_i, & \mathcal{B}_i&\rightarrow\mathcal{B}_i.
\end{align}
Thus, $\lambda_i$ is associated with the undetermined common boost, while $\mathcal{B}_i$ is gauge invariant and determines the physical relative boosts.

Substituting \eqref{boost_solution_general} back into the definition of $\mathcal{B}_i$ gives,
\begin{align}
\label{B_solution}
    \mathcal{B}_i=-\left[\sum_{\J}\frac{\beta^{\J}y_{\J}}{1-\widetilde{w}_{\J}}n_i^{\J}\right]\left[\sum_{\J}\frac{\beta^{\J}y_{\J}\widetilde{w}_{\J}}{1-\widetilde{w}_{\J}}\right]^{-1}.
\end{align}
The relative boosts are consequently,
\begin{align}
\label{boost_sol}
    A_i^{\I}-A_i^{\J}=\frac{1+\widetilde{w}_{\I}}{1-\widetilde{w}_{\I}}\left(n_i^{\I}+\mathcal{B}_i\right)-\frac{1+\widetilde{w}_{\J}}{1-\widetilde{w}_{\J}}\left(n_i^{\J}+\mathcal{B}_i\right),
\end{align}
with $\mathcal{B}_i$ given by \eqref{B_solution}.

The remaining overall Lorentz freedom may be fixed by choosing the gauge $\lambda_i=0$, or equivalently,
\begin{align}
    \sum_{\J}\beta^{\J}y_{\J}\left[A_i^{\J}+n_i^{\J}-\frac{1}{\widetilde{w}_{\J}}\left(A_i^{\J}-n_i^{\J}\right)\right]=0.
\end{align}
In this gauge, the individual boost vectors are directly given by,
\begin{align}
\label{boost_solution_gauge}
    A_i^{\I}=\frac{1+\widetilde{w}_{\I}}{1-\widetilde{w}_{\I}}\left(n_i^{\I}+\mathcal{B}_i\right).
\end{align}
For proportional backgrounds, $c_{\I}=1$ and $\widetilde{w}_{\I}=-1$, so the prefactor in \eqref{boost_solution_gauge} vanishes. All relative boosts therefore vanish, leaving only the redundant common Lorentz boost.

\subsection{Perturbative Bianchi constraints}

With the linear Lorentz constraints solved and $\omega^\I_{\mu\nu}$ eliminated, only the metric perturbations remain. We now proceed to the linear Bianchi constraints.

Expanding the interaction as,
\begin{align}
    V^{\mu}_{\I\,\nu}=\overline{V}{}^{\mu}_{\I\,\nu}+\delta V^{\mu}_{\I\,\nu},
\end{align}
the constraints take the form,
\begin{align}
    {}^{\I}\nabla_\mu V^{\mu}_{\I\,\nu}={}^{\I}\nabla_\mu\overline{V}^{\mu}_{\I\,\nu}+{}^{\I}\nabla_\mu\delta V^{\mu}_{\I\,\nu}=0.
\end{align}
Introducing the background connections ${}^\I\overline{\nabla}^{\,}_\mu \,\overline{g}^\I_{\alpha \beta}=0$, the Christoffel symbols are related by,
\begin{align}
    {}^{\I}\Gamma^\mu{}_{\!\alpha\beta}={}^{\I}\overline{\Gamma}{}^\mu{}_{\!\alpha\beta}+{}^{\I}C^\mu{}_{\!\alpha\beta},
\end{align}
where,
\begin{align}
\label{C_form}
    {}^{\I}C^\mu{}_{\!\alpha\beta}=\tfrac12\overline{g}^{\mu\nu}_{\I}\left({}^{\I}\overline{\nabla}_{\alpha}\delta g^{\I}_{\nu\beta}+{}^{\I}\overline{\nabla}_{\beta}\delta g^{\I}_{\alpha\nu}-{}^{\I}\overline{\nabla}_{\nu}\delta g^{\I}_{\alpha\beta}\right).
\end{align}
Imposing the background equations ${}^\I\overline{\nabla}_\mu \overline{V}{}^\mu_{\I\, \nu}=0$, the linearised Bianchi constraints become,
\begin{align}
\label{lin_bianchi}
    {}^{\I}\overline{\nabla}_\mu\delta V^{\mu}_{\I\,\nu}+{}^{\I}C^\mu{}_{\!\mu\sigma}\overline{V}{}^{\sigma}_{\I\,\nu}-{}^{\I}C^\sigma{}_{\!\mu\nu}\overline{V}{}^{\mu}_{\I\,\sigma}=0.
\end{align}
Using the effective-fluid variables \eqref{spin-2_w}, the temporal and spatial components become,
\begin{align}
    (\nu=0):\qquad
    &{}^{\I}\overline{\nabla}_\mu\delta V^{\mu}_{\I\,0}
    -\tfrac12\widetilde{\rho}_{\I}
    \left(1+\widetilde{w}_{\I}\right)
    {}^{\I}\overline{\nabla}_0\delta g^{i}_{\I\,i}=0,\\
    (\nu=i):\qquad
    &{}^{\I}\overline{\nabla}_\mu\delta V^{\mu}_{\I\,i}
    +\tfrac12\widetilde{\rho}_{\I}
    \left(1+\widetilde{w}_{\I}\right)
    {}^{\I}\overline{\nabla}_i\delta g^{0}_{\I\,0}=0.
\end{align}
For the background metrics \eqref{metric_ansatz}, the non-vanishing Christoffel symbols are,
\begin{align}
\label{christoffel_I}
    {}^{\I}\overline{\Gamma}^{0}{}_{00}&=\frac{\dot N_{\I}}{N_{\I}}, 
    & {}^{\I}\overline{\Gamma}^{0}{}_{ij}&=\frac{a_{\I}\dot a_{\I}}{N_{\I}^2}\gamma_{ij},\\
    {}^{\I}\overline{\Gamma}^{i}{}_{0j}&=\frac{\dot a_{\I}}{a_{\I}}\delta^i_j, 
    & {}^{\I}\overline{\Gamma}^{i}{}_{jk}&={}^{(\gamma)}\Gamma^{i}{}_{jk},
\end{align}
while ${}^{\I}\overline{\Gamma}^{0}{}_{0i}={}^{\I}\overline{\Gamma}^{i}{}_{00}=0$ and ${}^{(\gamma)}\Gamma^{i}{}_{jk}$ is the Levi–Civita connection of the spatial metric $\gamma_{ij}$,
\begin{align}
    {}^{(\gamma)}\Gamma^r{}_{rr}
    &=\frac{kr}{1-kr^2}, &
    {}^{(\gamma)}\Gamma^r{}_{\theta\theta}
    &=-r\left(1-kr^2\right), &
    {}^{(\gamma)}\Gamma^r{}_{\phi\phi}
    &=-r\left(1-kr^2\right)\sin^2\theta,\\
    {}^{(\gamma)}\Gamma^\theta{}_{r\theta}
    &={}^{(\gamma)}\Gamma^\theta{}_{\theta r}
    =\frac{1}{r}, &
    {}^{(\gamma)}\Gamma^\theta{}_{\phi\phi}
    &=-\sin\theta\cos\theta,\\
    {}^{(\gamma)}\Gamma^\phi{}_{r\phi}
    &={}^{(\gamma)}\Gamma^\phi{}_{\phi r}
    =\frac{1}{r}, &
    {}^{(\gamma)}\Gamma^\phi{}_{\theta\phi}
    &={}^{(\gamma)}\Gamma^\phi{}_{\phi\theta}
    =\cot\theta.
\end{align}
The corresponding covariant derivative is denoted by $D_i$, so that $D_i \gamma_{jk}=0$. 

Imposing the dynamical branch of the background Bianchi constraints, the Christoffel symbols \eqref{christoffel_I} can be written as,
\begin{align}
    {}^{\I}\overline{\Gamma}^{0}{}_{00}=H\frac{\mathrm{d}\ln\left(y_{\I}c_{\I}\right)}{\mathrm{d}\ln a}
    , && {}^{\I}\overline{\Gamma}^{0}{}_{ij}=\frac{a^2H}{c_{\I}}\gamma_{ij}, &&
    {}^{\I}\overline{\Gamma}^{i}{}_{0j}&=Hc_{\I}\delta^i_j.
\end{align}
The divergences of the interaction perturbations are consequently,
\begin{align}
\label{div_deltaV_0}
    {}^{\I}\overline{\nabla}_{\mu}\delta V^{\mu}_{\I\,0}&=\partial_t\delta V^{0}_{\I\,0}+D_i\delta V^{i}_{\I\,0}+Hc_{\I}\left(3\delta V^{0}_{\I\,0}-\delta V^{i}_{\I\,i}\right),\\
\label{div_deltaV_i}
    {}^{\I}\overline{\nabla}_{\mu}\delta V^{\mu}_{\I\,i}&=\partial_t\delta V^{0}_{\I\,i}+D_j\delta V^{j}_{\I\,i}+H\frac{\mathrm{d}\ln\left(a^2y_{\I}^3c_{\I}\right)}{\mathrm{d}\ln a}\delta V^{0}_{\I\,i}-\frac{a^2H}{c_{\I}}\gamma_{ij}\delta V^{j}_{\I\,0}.
\end{align}
To display the conservation-law structure of the constraints, we decompose the interaction perturbations into effective density, pressure, anisotropic stress and momentum perturbations,
\begin{align}
\label{spin2_fluid_perturbations}
    \delta V^0_{\I\,0}&=-\delta\widetilde{\rho}_{\I}, & \delta V^i_{\I\,j}&=\delta\widetilde{p}_{\I}\delta^i_j+\delta\widetilde{\pi}^i_{\I\,j}, & \delta\widetilde{\pi}^i_{\I\,i}&=0, & \delta\widetilde{q}^\I_{i}&=\frac{c_{\I}}{a}\delta V^0_{\I\,i}.
\end{align}
The remaining mixed component is not independent, since the covariant interaction tensor is symmetric. At linear order, $\delta V^{\I}_{0i}=\delta V^{\I}_{i0}$ gives,
\begin{align}
    -N_{\I}^2\delta V^0_{\I\,i}+N_{\I}a_{\I}\widetilde{p}_{\I}n_i^{\I}=a_{\I}^2\gamma_{ij}\delta V^j_{\I\,0}-N_{\I}a_{\I}\widetilde{\rho}_{\I}n_i^{\I}.
\end{align}
Using $a_{\I}=ay_{\I}$ and $N_{\I}=y_{\I}c_{\I}$, this becomes,
\begin{align}
\label{mixed_V_relation}
    \gamma_{ij}\delta V^j_{\I\,0}=\frac{c_{\I}}{a}\left[\left(\widetilde{\rho}_{\I}+\widetilde{p}_{\I}\right)n_i^{\I}-\delta\widetilde{q}^\I_{i}\right].
\end{align}
Before eliminating the boost perturbations $A^i_\I$, the effective momentum can be written as,
\begin{align}
    \delta\widetilde{q}^{\I}_{i}=\widetilde{\rho}_{\I}n_i^{\I}+\frac{\widetilde{\rho}_{\I}\widetilde{w}_{\I}}{1+\widetilde{w}_{\I}}\left(A_i^{\I}-\lambda_i\right),
\end{align}
so that it depends only on the boost relative to the undetermined common Lorentz boost. Substituting the Lorentz-constraint solution \eqref{boost_solution_general} then gives,
\begin{align}
\label{q_solution}
    \delta\widetilde{q}^{\I}_{i}=\frac{\widetilde{\rho}_{\I}}{1-\widetilde{w}_{\I}}\left(n_i^{\I}+\widetilde{w}_{\I}\mathcal{B}_i\right).
\end{align}
It is useful to introduce the momentum relative to the shift,
\begin{align}
\label{Q_solution}
    \delta\widetilde{\mathcal{Q}}_{\I\,i}&=\delta\widetilde{q}_{\I\,i}-\left(\widetilde{\rho}_{\I}+\widetilde{p}_{\I}\right)n_i^{\I}=\frac{\widetilde{\rho}_{\I}\widetilde{w}_{\I}}{1-\widetilde{w}_{\I}}\left(\mathcal{B}_i+\widetilde{w}_{\I}n_i^{\I}\right).
\end{align}
We first consider the temporal constraint. Substituting \eqref{spin2_fluid_perturbations} and \eqref{mixed_V_relation} into \eqref{div_deltaV_0} gives,
\begin{align}
    {}^{\I}\overline{\nabla}_{\mu}\delta V^{\mu}_{\I\,0}=-\partial_t\delta\widetilde{\rho}_{\I}-3Hc_{\I}\left(\delta\widetilde{\rho}_{\I}+\delta\widetilde{p}_{\I}\right)+\frac{c_{\I}}{a}D_i\left[\left(\widetilde{\rho}_{\I}+\widetilde{p}_{\I}\right)n_{\I}^i-\delta\widetilde{q}_{\I}^{\,i}\right].
\end{align}
The temporal Bianchi constraint therefore reduces to,
\begin{align}
\label{linear_bianchi_energy}
    \partial_t\delta\widetilde{\rho}_{\I}
    +3Hc_{\I}\left(
        \delta\widetilde{\rho}_{\I}
        +\delta\widetilde{p}_{\I}
    \right)
    +\frac{c_{\I}}{a}
    D_i\delta\widetilde{\mathcal{Q}}_{\I}^{\,i}
    +\tfrac12\left(
        \widetilde{\rho}_{\I}
        +\widetilde{p}_{\I}
    \right)
    \partial_t h^i_{\I\,i}
    =0.
\end{align}
This has the form of a regular perturbed continuity equation, except that $c_\I$ varies in time. 

For the spatial constraint, the definition of $\delta\widetilde{q}^\I_{i}$ and the background relation \eqref{c_def} imply,
\begin{align}
    \partial_t\delta V^0_{\I\,i}+H\frac{\mathrm{d}\ln\left(a^2y_{\I}^3c_{\I}\right)}{\mathrm{d}\ln a}\delta V^0_{\I\,i}=\frac{a}{c_{\I}}\left[\partial_t\delta\widetilde{q}^\I_{i}+3Hc_{\I}\delta\widetilde{q}^\I_{i}\right].
\end{align}
Using this result together with \eqref{mixed_V_relation}, the spatial divergence \eqref{div_deltaV_i} becomes,
\begin{align}
    {}^{\I}\overline{\nabla}_{\mu}\delta V^{\mu}_{\I\,i}=\frac{a}{c_{\I}}\left[\partial_t\delta\widetilde{q}^\I_{i}+4Hc_{\I}\delta\widetilde{q}^\I_{i}\right]+D_i\delta\widetilde{p}_{\I}+D_j\delta\widetilde{\pi}^j_{\I\,i}-aH\left(\widetilde{\rho}_{\I}+\widetilde{p}_{\I}\right)n_i^{\I}.
\end{align}
The last term cancels the shift contribution in ${}^{\I}\overline{\nabla}_i\delta g^0_{\I\,0}$. The spatial Bianchi constraint consequently takes the form,
\begin{align}
\label{linear_bianchi_momentum}
    \partial_t\delta\widetilde{q}^\I_{i}+4Hc_{\I}\delta\widetilde{q}^I_{i}+\frac{c_{\I}}{a}\left[D_i\delta\widetilde{p}_{\I}+D_j\delta\widetilde{\pi}^j_{\I\,i}+\left(\widetilde{\rho}_{\I}+\widetilde{p}_{\I}\right)D_i\phi_{\I}\right]=0.
\end{align}
Equations \eqref{linear_bianchi_energy} and \eqref{linear_bianchi_momentum} are the effective continuity and momentum constraints for each spin-2 sector. The momentum variables are not independent perturbative fields, since \eqref{q_solution} and \eqref{Q_solution} determine them algebraically through the metric shifts and the relative-boost variable $\mathcal{B}_i$. 

After substituting the algebraic expressions, the Bianchi constraints determine the relative lapse $\phi_\I$ and shift $n^i_\I$ perturbations. In particular, the lapse dependence of the effective pressure perturbation is,
\begin{align}
\label{pressure_lapse_dependence}
    \left.\delta\widetilde p_{\I}\right|_{\phi}
    &=\widetilde p_{\I}\Big[\frac{1}{S_N}\sum_{\J}\beta^{\J}y_{\J}c_{\J}\phi_{\J}-\phi_{\I}\Big].
\end{align}
The temporal Bianchi constraints \eqref{linear_bianchi_energy} therefore contain only relative lapse perturbations and determine $\Ncal-1$ of them algebraically. Once these have been eliminated, the spatial constraints \eqref{linear_bianchi_momentum} provide first-order equations for the $3(\Ncal-1)$ momenta.  The momentum can likewise be written entirely in terms of relative shift combinations. Using \eqref{q_solution}, \eqref{B_solution} and $\widetilde w_{\I}=\widetilde w/c_{\I}$ gives,
\begin{align}
\label{relative_shift_momentum}
    \delta\widetilde q_i^{\I}
    =\frac{\widetilde\rho_{\I}}{c_{\I}-\widetilde w}\bigg[\sum_{\J}\frac{\beta^{\J}y_{\J}}{c_{\J}-\widetilde w}\bigg]^{-1}\bigg[\sum_{\J}\frac{\beta^{\J}y_{\J}}{c_{\J}-\widetilde w}\left(c_{\I}n_i^{\I}-c_{\J}n_i^{\J}\right)\bigg].
\end{align}
Thus, the momentum depends only on differences between the weighted shifts $c_{\I}n_i^{\I}$. The spatial constraints thus provide first-order equations for the $3(\Ncal-1)$ relative shifts. The common lapse and shift remain undetermined, reflecting the surviving diagonal diffeomorphism invariance. 

\subsection{Perturbative field equations}

With the Lorentz constraints solved and the Bianchi constraints established, we now turn to the linearised field equations. Rotational symmetry of the homogeneous and isotropic backgrounds ensures that scalar, vector and tensor perturbations decouple at linear order. We derive the complete dynamical equations in the tensor sector, which contains no lapse, shift, or Lorentz perturbations. For the scalar and vector sectors, we restrict the analysis to the Lorentz and Bianchi constraints obtained above.

\subsubsection{Tensor perturbations}
\label{sec:tensor_perturbations}

We introduce the transverse-traceless spatial metric perturbations through,
\begin{align}
    \delta g^\I_{00}\big|_{\scriptscriptstyle\mathrm{TT}}=0, && 
    \delta g^\I_{0i}\big|_{\scriptscriptstyle\mathrm{TT}}=0, && 
    \delta g^\I_{ij}\big|_{\scriptscriptstyle\mathrm{TT}}=a_{\I}^2\mathfrak h^\I_{ij},
\end{align}
where,
\begin{align}
    D^i\mathfrak h^\I_{ij}=0, && \mathfrak h^{i}_{\I\,i}=0,
\end{align}
and spatial indices are raised and lowered using $\gamma_{ij}$. The linear Lorentz and Bianchi constraints vanish identically in this sector, so the tensor perturbations are determined directly by the transverse-traceless part of the field equations.

It is useful to introduce the weighted combination,
\begin{align}
    \mathfrak h^i_{u\,j}=\frac{1}{S_y}\sum_{\J}\beta^{\J}y_{\J}\mathfrak h^i_{\J\,j},
\end{align}
in terms of which the interaction perturbation becomes,
\begin{align}
\label{tensor_interaction}
    \left.\delta V^i_{\I\,j}\right|_{\scriptscriptstyle\mathrm{TT}}=\frac{\widetilde\rho_{\I}\widetilde w_{\I}}{2}\left(\mathfrak h^i_{\I\,j}-\mathfrak h^i_{u\,j}\right).
\end{align}
Thus, the interaction depends only on the difference between each tensor perturbation and the weighted average of all sectors.

For the background metric \eqref{metric_ansatz}, the transverse-traceless part of the linearised Einstein tensor is,
\begin{align}
\label{tensor_Einstein}
    \left.\delta G^i_{\I\,j}\right|_{\scriptscriptstyle\mathrm{TT}}=\frac{1}{2N_{\I}^2}\left[\ddot{\mathfrak h}^i_{\I\,j}+\left(3\frac{\dot a_{\I}}{a_{\I}}-\frac{\dot N_{\I}}{N_{\I}}\right)\dot{\mathfrak h}^i_{\I\,j}\right]-\frac{1}{2a_{\I}^2}\left(D^2-2k\right)\mathfrak h^i_{\I\,j}.
\end{align}
We retain a possible tensor anisotropic stress in the matter sector coupled to the first metric,
\begin{align}
    \left.\delta T^i_{1\,j}\right|_{\scriptscriptstyle\mathrm{TT}}=\delta\pi^i_j, &&
    D_i\delta\pi^i_j=0, && 
    \delta\pi^i_i=0,
\end{align}
which vanishes for a perfect fluid at linear order. The remaining matter sectors are absent, so $\delta T^i_{\I\,j}=0$ for $I\neq1$.

The tensor equations follow from the transverse-traceless part of the linearised field equations,
\begin{align}
    \left.\delta G^i_{\I\,j}\right|_{\scriptscriptstyle\mathrm{TT}}=\frac{1}{m_{\I}^2}\left[\left.\delta T^i_{\I\,j}\right|_{\scriptscriptstyle\mathrm{TT}}+\left.\delta V^i_{\I\,j}\right|_{\scriptscriptstyle\mathrm{TT}}\right].
\end{align}
For $I=1$, the equation reads,
\begin{align}
\label{tensor_eom_1}
    \ddot{\mathfrak h}^i_{1\,j}+3H\dot{\mathfrak h}^i_{1\,j}-\frac{1}{a^2}\left(D^2-2k\right)\mathfrak h^i_{1\,j}=\frac{2}{m_1^2}\delta\pi^i_j+\frac{\widetilde p}{m_1^2}\left(\mathfrak h^i_{1\,j}-\mathfrak h^i_{u\,j}\right).
\end{align}
For $I\neq1$, using the dynamical-branch relations $a_{\I}=ay_{\I}$ and $N_{\I}=y_{\I}c_{\I}$, the equations take the form,
\begin{align}
\label{tensor_eom_I}
    \ddot{\mathfrak h}^i_{\I\,j}+\left[H\left(2c_{\I}+1\right)-\frac{\dot c_{\I}}{c_{\I}}\right]\dot{\mathfrak h}^i_{\I\,j}-\frac{c_{\I}^2}{a^2}\left(D^2-2k\right)\mathfrak h^i_{\I\,j}=\frac{\alpha_{\I}c_{\I}}{m_1^2y_{\I}}\widetilde p\left(\mathfrak h^i_{\I\,j}-\mathfrak h^i_{u\,j}\right).
\end{align}
To display the coupled structure, we collect the perturbations into,
\begin{align}
    \boldsymbol{\mathfrak h}^{\,i}{}_{j}=\left(\mathfrak h^i_{1\,j},\mathfrak h^i_{2\,j},\ldots,\mathfrak h^i_{\scriptscriptstyle{\Ncal}\,j}\right)^{\T},
\end{align}
and define,
\begin{align}
    r_{\I}=\frac{\beta^{\I}y_{\I}}{S_y}, && \sum_{\I}r_{\I}=1, && \mathfrak h^i_{u\,j}=\sum_{\J}r_{\J}\mathfrak h^i_{\J\,j}.
\end{align}
We also introduce the interaction coefficients,
\begin{align}
    \mathfrak m_1^2=-\frac{\widetilde p}{m_1^2}, && \mathfrak m_{\I}^2=-\frac{\alpha_{\I}c_{\I}}{m_1^2y_{\I}}\widetilde p, \qquad I\neq1,
\end{align}
and the diagonal propagation and damping matrices,
\begin{align}
    \mathsf C^2_{\I\J}=c_{\I}^2\delta_{\I\J}, && \mathsf F_{\I\J}=F_{\I}\delta_{\I\J},
\end{align}
where,
\begin{align}
    F_1=3H, && F_{\I}=H\left(2c_{\I}+1\right)-\frac{\dot c_{\I}}{c_{\I}}, \qquad I\neq1.
\end{align}
The interaction matrix is then,
\begin{align}
\label{tensor_mass_matrix}
    \left(\mathsf M^2\right)_{\I\J}=\mathfrak m_{\I}^2\left(\delta_{\I\J}-r_{\J}\right),
\end{align}
and the complete tensor system takes the form,
\begin{align}
\label{tensor_matrix_equation}
    \ddot{\boldsymbol{\mathfrak h}}^{\,i}{}_{j}+\mathsf F\dot{\boldsymbol{\mathfrak h}}^{\,i}{}_{j}-\frac{1}{a^2}\mathsf C^2\left(D^2-2k\right)\boldsymbol{\mathfrak h}^{\,i}{}_{j}+\mathsf M^2\boldsymbol{\mathfrak h}^{\,i}{}_{j}=\boldsymbol{\mathcal S}^{\,i}{}_{j},
\end{align}
with,
\begin{align}
    \boldsymbol{\mathcal S}^{\,i}{}_{j}=\left(\tfrac{2}{m_1^2}\delta\pi^i_j,0,\ldots,0\right)^{\T}.
\end{align}
If the tensor anisotropic stress vanishes, the equations form a homogeneous linear system and we can expand the perturbations in transverse-traceless tensor harmonics,
\begin{align}
    \boldsymbol{\mathfrak h}^{\,i}{}_{j}(t,\boldsymbol x)=\sum_s\int\mathrm d q_{\T}\,\boldsymbol{\mathfrak h}_s(t,q_{\T})Q^i_{s\,j}(\boldsymbol x,q_{\T}),
\end{align}
where $s=1,2$ labels the two tensor polarisations, and the harmonics satisfy,
\begin{align}
    -\left(D^2-2k\right)Q^i_{s\,j}=q_{\T}^2Q^i_{s\,j}.
\end{align}
Since every harmonic evolves independently, we suppress $s$ and $q_{\T}$ and obtain,
\begin{align}
\label{tensor_mode_matrix_equation}
    \ddot{\boldsymbol{\mathfrak h}}+\mathsf F\dot{\boldsymbol{\mathfrak h}}+\Big[\frac{q_{\T}^2}{a^2}\mathsf C^2+\mathsf M^2\Big]\boldsymbol{\mathfrak h}=0.
\end{align}
The rows of $\mathsf M^2$ sum to zero,
\begin{align}
    \mathsf M^2\left(1,1,\ldots,1\right)^{\T}=0,
\end{align}
so a common tensor perturbation is unaffected by the interaction. On a general cosmological background, however, this does not define a decoupled massless mode because the damping, propagation and interaction matrices cannot generically be diagonalised simultaneously. In particular, since $\mathsf C^2$ and $\mathsf M^2$ are not generally simultaneously diagonalisable, the propagation eigenmodes and mass eigenmodes are different. A perturbation generated in one sector may therefore be a superposition of modes that accumulate different phases during propagation, potentially giving rise to neutrino-oscillation-like behaviour.

The mass eigenstates become well defined on a proportional vacuum background, where $y_{\I}=\bar y_{\I}$ and $c_{\I}=1$. In this case,
\begin{align}
    \mathsf C^2=\id, &&\mathsf F=3H\id,
\end{align}
and \eqref{tensor_mode_matrix_equation} reduces to,
\begin{align}
    \ddot{\boldsymbol{\mathfrak h}}+3H\dot{\boldsymbol{\mathfrak h}}+\Big[\frac{q_{\T}^2}{a^2}\id+\mathsf M^2\Big]\boldsymbol{\mathfrak h}=0.
\end{align}
The different kinetic normalisations are removed by defining,
\begin{align}
    \widehat{\mathfrak h}_{\I}=m_{\I}\bar y_{\I}\mathfrak h_{\I}, && \widehat{\boldsymbol{\mathfrak h}}=\mathsf K^{1/2}\boldsymbol{\mathfrak h}, && \mathsf K_{\I\J}=m_{\I}^2\bar y_{\I}^2\delta_{\I\J}.
\end{align}
The mass matrix in this basis is,
\begin{align}
    \widehat{\mathsf M}^2=\mathsf K^{1/2}\mathsf M^2\mathsf K^{-1/2},
\end{align}
which is symmetric,
\begin{align}
    \left(\mathsf K\mathsf M^2\right)_{\I\J}=m^4\overline S_y^4\left(r_{\I}\delta_{\I\J}-r_{\I}r_{\J}\right).
\end{align}
It can therefore be diagonalised by an orthogonal matrix $\mathsf O$,
\begin{align}
    \mathsf O^{\T}\widehat{\mathsf M}^2\mathsf O=\operatorname{diag}\left(0,\mu_1^2,\ldots,\mu_{{\scriptscriptstyle \Ncal}-1}^2\right),
\end{align}
where the $\mu_n^2$ are the physical Fierz–Pauli eigenmasses obtained in \cite{Flinckman:2024zpb}. Defining,
\begin{align}
    \boldsymbol h=\mathsf O^{\T}\widehat{\boldsymbol{\mathfrak h}}=\left(h_0,h_1,\ldots,h_{{\scriptscriptstyle \Ncal}-1}\right)^{\T},
\end{align}
gives the decoupled equations,
\begin{align}
    \ddot h_0+3H\dot h_0+\frac{q_{\T}^2}{a^2}h_0&=0,\\
    \ddot h_n+3H\dot h_n+\Big[\frac{q_{\T}^2}{a^2}+\mu_n^2\Big]h_n&=0, \qquad n=1,\ldots,\Ncal-1.
\end{align}
The massless combination is the kinetically weighted common perturbation,
\begin{align}
    h_0\propto\sum_{\I}m_{\I}^2\bar y_{\I}^2\mathfrak h_{\I}.
\end{align}
The variables $h_n$ describe only the helicity-$2$ components of the corresponding mass eigenfields. However, for proportional backgrounds the same transformation diagonalises the covariant quadratic action into one massless and $\Ncal-1$ massive Fierz–Pauli actions \cite{Flinckman:2024zpb}. All helicities of a given mass eigenfield therefore share the same characteristic cone and have characteristic speed unity relative to the matter-coupled metric.

Locally, in the WKB regime, the physical momentum can be approximated as $p=q_{\T}/a$ and the dispersion relations are,
\begin{align}
    \omega_0^2=p^2, \qquad \omega_n^2=p^2+\mu_n^2.
\end{align}
The massive wave packets consequently have group velocities,
\begin{align}
    v^{\mathrm g}_n=\frac{\partial\omega_n}{\partial p}=\frac{p}{\sqrt{p^2+\mu_n^2}}<1,
\end{align}
despite sharing the luminal characteristic speed.

Since matter couples to the first metric rather than directly to a mass eigenfield, a gravitational wave is generically produced and detected as a superposition of the massless and massive modes. While the wave packets remain coherent, their relative phases can produce gravitational-wave oscillations. If they separate during propagation, the slower massive components may be detected as delayed secondary signals, or gravitational-wave echoes \cite{Max:2017flc,Max:2017kdc,Cembranos:2026cqk}.

\subsubsection{Vector perturbations}
\label{sec:vector_perturbations}

We next consider the vector sector. Decomposing the shift and spatial metric perturbations into their transverse parts, we write,
\begin{align}
    n_i^{\I}=D_i n^{\I}+\mathfrak n_i^{\I}, \qquad \left.h_{ij}^{\I}\right|_{\scriptscriptstyle\mathrm V}=2D_{(i}E^{\I}_{j)}, \qquad D^i\mathfrak n_i^{\I}=D^iE_i^{\I}=0.
\end{align}
The vector metric perturbations are therefore,
\begin{align}
\label{vector_metric_perturbations}
    \left.\delta g^\I_{00}\right|_{\scriptscriptstyle\mathrm V}=0, \qquad \left.\delta g^\I_{0i}\right|_{\scriptscriptstyle\mathrm V}=N_{\I}a_{\I}\mathfrak n_i^{\I}, \qquad \left.\delta g^\I_{ij}\right|_{\scriptscriptstyle\mathrm V}=2a_{\I}^2D_{(i}E^{\I}_{j)}.
\end{align}
There is one common transverse spatial diffeomorphism, so that only relative combinations of the vectors $E_i^{\I}$ are independent.

Similarly, we decompose the Lorentz boost perturbations as,
\begin{align}
    A_i^{\I}=D_iA^{\I}+\mathcal A_i^{\I}, \qquad D^i\mathcal A_i^{\I}=0.
\end{align}
Since we have the general solution of the perturbative Lorentz constraints, the vector sector solution can easily be obtained by the transverse projection. The projections of \eqref{boost_solution_general} and \eqref{B_solution} gives,
\begin{align}
\label{vector_boost_solution}
    \mathcal A_i^{\I}=\mathfrak l_i+\frac{1+\widetilde w_{\I}}{1-\widetilde w_{\I}}\left(\mathfrak n_i^{\I}+\mathfrak b_i\right), \qquad \mathfrak b_i=-\left[\sum_{\J}\frac{\beta^{\J}y_{\J}}{1-\widetilde w_{\J}}\mathfrak n_i^{\J}\right]\left[\sum_{\J}\frac{\beta^{\J}y_{\J}\widetilde w_{\J}}{1-\widetilde w_{\J}}\right]^{-1}.
\end{align}
Here $\mathfrak l_i$ is the transverse part of the undetermined common Lorentz boost. It cancels from all relative boosts and may be removed using the diagonal local Lorentz symmetry. The antisymmetric spatial Lorentz perturbations are likewise common to all vielbeins and drop out of the field equations. The Lorentz constraints therefore determine the $\Ncal-1$ relative transverse boosts algebraically in terms of the transverse shifts.

We next consider the linear Bianchi constraints. In the vector sector, the effective density and isotropic pressure perturbations vanish, while the transverse interaction momentum obtained from \eqref{q_solution} is,
\begin{align}
\label{vector_interaction_momentum}
    \left.\delta\widetilde q^{\I}_{i}\right|_{\scriptscriptstyle\mathrm V}=\frac{\widetilde\rho_{\I}}{1-\widetilde w_{\I}}\left(\mathfrak n_i^{\I}+\widetilde w_{\I}\mathfrak b_i\right).
\end{align}
Introducing the weighted spatial vector,
\begin{align}
\label{weighted_vector}
    E_i^u=\frac{1}{S_y}\sum_{\J}\beta^{\J}y_{\J}E_i^{\J},
\end{align}
the vector part of the interaction anisotropic stress is,
\begin{align}
\label{vector_interaction_stress}
    \left.\delta\widetilde\pi^i_{\I\,j}\right|_{\scriptscriptstyle\mathrm V}=\widetilde\rho_{\I}\widetilde w_{\I}\gamma^{ik}D_{(k}\left(E^{\I}_{j)}-E^u_{j)}\right).
\end{align}
The temporal Bianchi constraints are identically satisfied in the vector sector, since the vector perturbations are transverse and the spatial metric perturbations are traceless. The transverse part of the spatial Bianchi constraints reduces to,
\begin{align}
\label{vector_bianchi}
    \partial_t\left.\delta\widetilde q^{\I}_{i}\right|_{\scriptscriptstyle\mathrm V}+4Hc_{\I}\left.\delta\widetilde q^{\I}_{i}\right|_{\scriptscriptstyle\mathrm V}+\frac{c_{\I}\widetilde\rho_{\I}\widetilde w_{\I}}{2a}\left(D^2+2k\right)\left(E_i^{\I}-E_i^u\right)=0.
\end{align}
Together with \eqref{vector_interaction_momentum}, these form first-order relations between the transverse shifts and the relative spatial metric vectors. Only $\Ncal-1$ of the vector Bianchi constraints are independent, since the remaining combination follows from diagonal diffeomorphism invariance once the Lorentz constraints have been imposed.

Thus, the Lorentz and Bianchi constraints determine how the non-dynamical boost and shift perturbations are related to the relative spatial vector perturbations $E^\I_i$. A complete derivation of the dynamical vector field equations and their stability properties is left for future work.

\subsubsection{Scalar perturbations}
\label{sec:scalar_perturbations}

We finally consider the scalar sector. Decomposing the shift and spatial metric perturbations into scalar, vector and tensor parts, we write,
\begin{align}
    n_i^{\I}=D_i n^{\I}+\mathfrak n_i^{\I}, \qquad \left.h_{ij}^{\I}\right|_{\scriptscriptstyle\mathrm S}=2\psi_{\I}\gamma_{ij}+2D_iD_jE_{\I}.
\end{align}
The scalar metric perturbations are therefore,
\begin{align}
\label{scalar_metric_perturbations}
    \left.\delta g^\I_{00}\right|_{\scriptscriptstyle\mathrm S}=-2N_{\I}^2\phi_{\I}, \qquad \left.\delta g^\I_{0i}\right|_{\scriptscriptstyle\mathrm S}=N_{\I}a_{\I}D_i n^{\I}, \qquad \left.\delta g^\I_{ij}\right|_{\scriptscriptstyle\mathrm S}=2a_{\I}^2\left(\psi_{\I}\gamma_{ij}+D_iD_jE_{\I}\right).
\end{align}
In particular, the trace of the spatial metric perturbation is,
\begin{align}
\label{scalar_metric_trace}
    h^i_{\I\,i}=6\psi_{\I}+2D^2E_{\I}.
\end{align}
There are two common scalar diffeomorphisms, corresponding to a time reparametrisation and a longitudinal spatial diffeomorphism. Consequently, only relative combinations of the scalar perturbations describe the additional gravitational degrees of freedom.

The scalar part of the Lorentz boost is introduced through,
\begin{align}
    A_i^{\I}=D_iA^{\I}+\mathcal A_i^{\I}.
\end{align}
Similarly, we decompose the quantities entering the general Lorentz-constraint solution as $\lambda_i=D_i\lambda+\mathfrak l_i$ and $\mathcal B_i=D_i\mathcal B+\mathfrak b_i$. The longitudinal part of \eqref{boost_solution_general} then gives,
\begin{align}
\label{scalar_boost_solution}
    A^{\I}=\lambda+\frac{1+\widetilde w_{\I}}{1-\widetilde w_{\I}}\left(n^{\I}+\mathcal B\right), \qquad \mathcal B=-\left[\sum_{\J}\frac{\beta^{\J}y_{\J}}{1-\widetilde w_{\J}}n^{\J}\right]\left[\sum_{\J}\frac{\beta^{\J}y_{\J}\widetilde w_{\J}}{1-\widetilde w_{\J}}\right]^{-1}.
\end{align}
Here $\lambda$ represents the undetermined common Lorentz boost and may be removed using the diagonal local Lorentz symmetry. The Lorentz constraints therefore determine the $\Ncal-1$ relative scalar boosts $A^{\I}-A^{\J}$ algebraically in terms of the scalar shifts $n^{\I}$. The antisymmetric spatial Lorentz perturbations do not contribute to the scalar sector.

To obtain the scalar Bianchi constraints, we decompose the interaction momentum and anisotropic stress as,
\begin{align}
    \delta\widetilde q_i^{\I}\Big|_{\scriptscriptstyle\mathrm S}=D_i\delta\widetilde q_{\I}, 
    \qquad \delta\widetilde{\mathcal Q}_{\I\,i}\Big|_{\scriptscriptstyle\mathrm S}=D_i\delta\widetilde{\mathcal Q}_{\I}, 
    \qquad \delta\widetilde\pi^i_{\I\,j}\Big|_{\scriptscriptstyle\mathrm S}=\left(D^iD_j-\tfrac{1}{3}\delta^i_jD^2\right)\delta\widetilde\pi_{\I}.
\end{align}
The scalar interaction momenta are determined by the Lorentz-constraint solution,
\begin{align}
\label{scalar_interaction_momentum}
    \delta\widetilde q_{\I}=\frac{\widetilde\rho_{\I}}{1-\widetilde w_{\I}}\left(n^{\I}+\widetilde w_{\I}\mathcal B\right), \qquad \delta\widetilde{\mathcal Q}_{\I}=\frac{\widetilde\rho_{\I}\widetilde w_{\I}}{1-\widetilde w_{\I}}\left(\mathcal B+\widetilde w_{\I}n^{\I}\right).
\end{align}
The effective density, pressure and scalar anisotropic-stress perturbations $\delta\widetilde\rho_{\I}$, $\delta\widetilde p_{\I}$ and $\delta\widetilde\pi_{\I}$ are obtained by taking the corresponding scalar projections of the interaction perturbations derived above.

The temporal Bianchi constraints become,
\begin{align}
\label{scalar_bianchi_energy}
    \partial_t\delta\widetilde\rho_{\I}
    +3Hc_{\I}\left(
        \delta\widetilde\rho_{\I}
        +\delta\widetilde p_{\I}
    \right)
    +\frac{c_{\I}}{a}D^2\delta\widetilde{\mathcal Q}_{\I}
    +\left(
        \widetilde\rho_{\I}
        +\widetilde p_{\I}
    \right)
    \partial_t\left(
        3\psi_{\I}+D^2E_{\I}
    \right)
    =0.
\end{align}
For the scalar anisotropic stress defined above, its divergence satisfies,
\begin{align}
    D_j\left.\delta\widetilde\pi^j_{\I\,i}\right|_{\scriptscriptstyle\mathrm S}=\tfrac{2}{3}D_i\left(D^2+3k\right)\delta\widetilde\pi_{\I}.
\end{align}
The longitudinal part of the spatial Bianchi constraints therefore reduces to,
\begin{align}
\label{scalar_bianchi_momentum}
    \partial_t\delta\widetilde q_{\I}+4Hc_{\I}\delta\widetilde q_{\I}+\frac{c_{\I}}{a}\left[\delta\widetilde p_{\I}+\tfrac{2}{3}\left(D^2+3k\right)\delta\widetilde\pi_{\I}+\left(\widetilde\rho_{\I}+\widetilde p_{\I}\right)\phi_{\I}\right]=0.
\end{align}
Only $2(\Ncal-1)$ of the scalar Bianchi constraints \eqref{scalar_bianchi_energy} and \eqref{scalar_bianchi_momentum} are independent, since one temporal and one longitudinal combination follow from diagonal diffeomorphism invariance once the Lorentz constraints have been imposed.

Thus, the scalar Lorentz constraints eliminate the relative boost scalars, while the temporal and longitudinal Bianchi constraints provide first-order relations among the relative lapse, shift and spatial metric perturbations. A complete derivation of the dynamical scalar field equations and their stability and phenomenological consequences is left for future work.

\section{Summary and outlook}

In this work, we have initiated the study of homogeneous and isotropic cosmologies in the ghost-free Hassan--Schmidt-May multi-gravity theory. The background Lorentz constraints eliminate the relative Lorentz fields, while the Bianchi constraints divide the solutions into an algebraic and a dynamical branch. On the dynamical branch, the cosmological equations reduce to a modified Friedmann equation together with $\Ncal-1$ coupled quartic equations for the scale-factor ratios. We showed that these equations admit at least one solution with all scale factors positive for every positive matter energy density when the interaction parameters $\beta^{\I}$ have the same sign.

The interaction contribution was expressed as an effective spin-2 fluid. We identified a regular high-density branch approaching the standard matter- or radiation-dominated evolution and determined the low-density expansion around proportional vacuum solutions. The constant part of the interaction acts as dark energy, while the leading matter-dependent correction can mimic an additional pressureless component at the level of the background evolution. Together with the spectrum of massive spin-2 fields, this raises the possibility that light and heavy sectors could contribute differently to dark energy and dark matter. The modified evolution between the high- and low-density limits also provides additional freedom in the expansion history and may therefore be relevant to the Hubble tension, as has recently been investigated in bimetric cosmology \cite{Hogas:2025ahb}.

The null cones of the metrics are coaxial on the dynamical branch, approach fixed relative openings at high density and coincide in the vacuum limit. Their larger opening angles in the early Universe enlarge the causal domains associated with the additional metrics and could therefore be relevant to the horizon problem underlying the observed homogeneity of the cosmic microwave background. Since Standard Model matter couples only to the first metric, however, establishing whether this enlarged causal structure affects the thermal history of the matter sector requires a more complete analysis.

At the perturbative level, we derived and solved the linear Lorentz constraints, obtained the linear Bianchi constraints and their scalar and vector projections, and derived the complete coupled tensor perturbation equations. On proportional vacuum backgrounds, the tensor system diagonalises into one massless and $\Ncal-1$ massive tensor modes, allowing for gravitational-wave oscillations and delayed massive signals. A complete stability analysis of the scalar and vector sectors remains an important next step.

The phenomenological viability of these cosmologies will also depend on their nonlinear behaviour. In bimetric gravity, the Vainshtein mechanism \cite{Vainshtein:1972sx} restores general relativity near sufficiently dense sources \cite{Babichev:2013pfa}, while related nonlinear screening has also been argued to operate for cosmological perturbations \cite{Mortsell:2015exa,Luben:2019yyx}. Whether an analogous mechanism exists for genuine multi-field interactions, potentially with several characteristic scales associated with the different spin-2 masses, remains to be established. The framework developed here provides the starting point for addressing these questions and for confronting multi-gravity cosmology with observations.


\newpage
\appendix

\section{Existence of positive scale-factor solutions}
\label{app:positive_cosmological_solutions}

We now show that the cosmological consistency equations \eqref{friedmann_poly} admit at least one solution with positive scale-factor ratios for every fixed matter density $\rho\geq0$, provided that all interaction parameters $\beta^{\I}$ are non-zero and have the same sign. The argument closely follows the proof for proportional backgrounds given in \cite{Flinckman:2024zpb}.

For $I\neq1$, we introduce the rescaled variables,
\begin{align}
\label{positive_z_definition}
    z_{\I}=\frac{\beta^{\I}}{\beta^1}y_{\I}.
\end{align}
Since all $\beta^{\I}$ have the same sign, $z_{\I}>0$ is equivalent to $y_{\I}>0$. Using $y_1=1$, the weighted scale-factor sum and the interaction energy density become,
\begin{align}
    S_y=\beta^1\Big[1+\sum_{\J\neq1}z_{\J}\Big], \qquad \widetilde\rho=m^4\left(\beta^1\right)^4\Big[1+\sum_{\J\neq1}z_{\J}\Big]^3.
\end{align}
Multiplying each polynomial \eqref{friedmann_poly} by the positive factor $\beta^{\I}/\beta^1$, we define the equivalent polynomials $\widehat{\mathcal P}_{\I}=\beta^{\I}\mathcal P_{\I}/\beta^1$,
\begin{align}
\label{positive_polynomials}
    \widehat{\mathcal P}_{\I} =
    m_1^2\Lambda_{\I}\left(\tfrac{\beta^1}{\beta^{\I}}\right)^2z_{\I}^3
    -\left(m_1^2\Lambda_1+\rho\right)z_{\I}
    +m^4\left(\beta^1\right)^4\Big[
        \tfrac{m_1^2}{m_{\I}^2}
        \left(\tfrac{\beta^{\I}}{\beta^1}\right)^2
        -z_{\I}
    \Big]
    \Big[1+\sum_{\J\neq1}z_{\J}\Big]^3.
\end{align}
Consider the hypercube $0\leq z_{\I}\leq M$ for all $I\neq1$. On the lower face $z_{\I}=0$, the corresponding polynomial satisfies,
\begin{align}
    \left.\widehat{\mathcal P}_{\I}\right|_{z_{\I}=0}
    =
    \tfrac{m_1^2}{m_{\I}^2}
    \left(\tfrac{\beta^{\I}}{\beta^1}\right)^2
    m^4\left(\beta^1\right)^4
    \Big[1+\sum_{\substack{\J\neq1\\\J\neq\I}}z_{\J}\Big]^3
    >0,
\end{align}
since $z_{J\neq I}\geq0$. 

On the opposite face $z_{\I}=M$, the polynomial is,
\begin{align}
    \left.\widehat{\mathcal P}_{\I}\right|_{z_{\I}=M}
    &=
    m_1^2\Lambda_{\I}\left(\tfrac{\beta^1}{\beta^{\I}}\right)^2M^3
    -\left(m_1^2\Lambda_1+\rho\right)M \notag\\
    &\quad
    +m^4\left(\beta^1\right)^4
    \Big[
        \tfrac{m_1^2}{m_{\I}^2}
        \left(\tfrac{\beta^{\I}}{\beta^1}\right)^2-M
    \Big]
    \Big[
        1+M+\sum_{\substack{\J\neq1\\\J\neq\I}}z_{\J}
    \Big]^3.
\end{align}
Choose $M$ sufficiently large so that
$M>2(m_1^2/m_{\I}^2)(\beta^{\I}/\beta^1)^2$. Then,
\begin{align}
    \frac{m_1^2}{m_{\I}^2}
    \left(\frac{\beta^{\I}}{\beta^1}\right)^2-M
    <-\frac{M}{2},
    \qquad
    \Big[
        1+M+\sum_{\substack{\J\neq1\\\J\neq\I}}z_{\J}
    \Big]^3
    \geq M^3,
\end{align}
for every $z_{\J}\in[0,M]$. The interaction term is therefore bounded above by
$-\tfrac12m^4(\beta^1)^4M^4$ uniformly on the entire upper face. Consequently,
\begin{align}
    \left.\widehat{\mathcal P}_{\I}\right|_{z_{\I}=M}
    &\leq
    m_1^2\left|\Lambda_{\I}\right|
    \left(\tfrac{\beta^1}{\beta^{\I}}\right)^2M^3
    +\left|m_1^2\Lambda_1+\rho\right|M
    -\frac{1}{2}m^4\left(\beta^1\right)^4M^4
    <0,
\end{align}
for sufficiently large $M$, uniformly for all remaining
$z_{\J}\in[0,M]$. Since there are only finitely many indices, $M$ can be
chosen large enough that this holds for every $I\neq1$.

The polynomials $\widehat{\mathcal P}_{\I}$ therefore have opposite signs
on each pair of faces perpendicular to the $z_{\I}$ direction. The
Poincaré--Miranda theorem then guarantees the existence of a point in the
hypercube at which all $\widehat{\mathcal P}_{\I}$ vanish
simultaneously.\footnote{The Poincaré--Miranda theorem states that if $n$
continuous functions
$F_i:[a_1,b_1]\times\cdots\times[a_n,b_n]\rightarrow\mathbb{R}$ have
opposite signs on the two faces $x_i=a_i$ and $x_i=b_i$ for every $i$,
then there exists a point in the hyperrectangle at which all $F_i$
vanish simultaneously.} Since none of the polynomials vanish on the
lower or upper faces, the solution lies in the interior,
\begin{align}
    0<z_{\I}<M.
\end{align}
Finally, since $\beta^{\I}/\beta^1>0$, we obtain,
\begin{align}
    y_{\I}=\frac{\beta^1}{\beta^{\I}}z_{\I}>0.
\end{align}
Thus, for every fixed $\rho\geq0$, the cosmological consistency equations
admit at least one solution with all scale-factor ratios positive.

\bibliographystyle{unsrtnat}
\bibliography{ref.bib}

@article{Boulware:1972yco,
    author = "Boulware, D. G. and Deser, Stanley",
    title = "{Can gravitation have a finite range?}",
    doi = "10.1103/PhysRevD.6.3368",
    journal = "Phys. Rev. D",
    volume = "6",
    pages = "3368--3382",
    year = "1972"
}

@article{Boulware:1972zf,
    author = "Boulware, D. G. and Deser, Stanley",
    title = "{Inconsistency of finite range gravitation}",
    doi = "10.1016/0370-2693(72)90418-2",
    journal = "Phys. Lett. B",
    volume = "40",
    pages = "227--229",
    year = "1972"
}

@article{Creminelli:2005qk,
    author = "Creminelli, Paolo and Nicolis, Alberto and Papucci, Michele and Trincherini, Enrico",
    title = "{Ghosts in massive gravity}",
    eprint = "hep-th/0505147",
    archivePrefix = "arXiv",
    reportNumber = "HUTP-05-A0020, HD-THEP-05-09, UCB-PTH-05-14, LBNL-57558",
    doi = "10.1088/1126-6708/2005/09/003",
    journal = "JHEP",
    volume = "09",
    pages = "003",
    year = "2005"
}

@article{deRham:2010ik,
    author = "de Rham, Claudia and Gabadadze, Gregory",
    title = "{Generalization of the Fierz-Pauli Action}",
    eprint = "1007.0443",
    archivePrefix = "arXiv",
    primaryClass = "hep-th",
    reportNumber = "NYU-TH-06-13-10",
    doi = "10.1103/PhysRevD.82.044020",
    journal = "Phys. Rev. D",
    volume = "82",
    pages = "044020",
    year = "2010"
}

@article{deRham:2010kj,
    author = "de Rham, Claudia and Gabadadze, Gregory and Tolley, Andrew J.",
    title = "{Resummation of Massive Gravity}",
    eprint = "1011.1232",
    archivePrefix = "arXiv",
    primaryClass = "hep-th",
    doi = "10.1103/PhysRevLett.106.231101",
    journal = "Phys. Rev. Lett.",
    volume = "106",
    pages = "231101",
    year = "2011"
}

@article{Hassan:2011hr,
    author = "Hassan, S. F. and Rosen, Rachel A.",
    title = "{Resolving the Ghost Problem in non-Linear Massive Gravity}",
    eprint = "1106.3344",
    archivePrefix = "arXiv",
    primaryClass = "hep-th",
    doi = "10.1103/PhysRevLett.108.041101",
    journal = "Phys. Rev. Lett.",
    volume = "108",
    pages = "041101",
    year = "2012"
}

@article{Hassan:2011tf,
    author = "Hassan, S. F. and Rosen, Rachel A. and Schmidt-May, Angnis",
    title = "{Ghost-free Massive Gravity with a General Reference Metric}",
    eprint = "1109.3230",
    archivePrefix = "arXiv",
    primaryClass = "hep-th",
    doi = "10.1007/JHEP02(2012)026",
    journal = "JHEP",
    volume = "02",
    pages = "026",
    year = "2012"
}

@article{Hassan:2011ea,
    author = "Hassan, S. F. and Rosen, Rachel A.",
    title = "{Confirmation of the Secondary Constraint and Absence of Ghost in Massive Gravity and Bimetric Gravity}",
    eprint = "1111.2070",
    archivePrefix = "arXiv",
    primaryClass = "hep-th",
    doi = "10.1007/JHEP04(2012)123",
    journal = "JHEP",
    volume = "04",
    pages = "123",
    year = "2012"
}

@article{Hassan:2012qv,
    author = "Hassan, S. F. and Schmidt-May, Angnis and von Strauss, Mikael",
    title = {{Proof of Consistency of Nonlinear Massive Gravity in the St{\"u}ckelberg Formulation}},
    eprint = "1203.5283",
    archivePrefix = "arXiv",
    primaryClass = "hep-th",
    doi = "10.1016/j.physletb.2012.07.018",
    journal = "Phys. Lett. B",
    volume = "715",
    pages = "335--339",
    year = "2012"
}

@article{Comelli:2013txa,
    author = "Comelli, Denis and Nesti, Fabrizio and Pilo, Luigi",
    title = "{Massive gravity: a General Analysis}",
    eprint = "1305.0236",
    archivePrefix = "arXiv",
    primaryClass = "hep-th",
    doi = "10.1007/JHEP07(2013)161",
    journal = "JHEP",
    volume = "07",
    pages = "161",
    year = "2013"
}

@article{Hassan:2018mbl,
    author = "Hassan, S. F. and Lundkvist, Anders",
    title = "{Analysis of constraints and their algebra in bimetric theory}",
    eprint = "1802.07267",
    archivePrefix = "arXiv",
    primaryClass = "hep-th",
    doi = "10.1007/JHEP08(2018)182",
    journal = "JHEP",
    volume = "08",
    pages = "182",
    year = "2018"
}

@article{Hassan:2011zd,
    author = "Hassan, S. F. and Rosen, Rachel A.",
    title = "{Bimetric Gravity from Ghost-free Massive Gravity}",
    eprint = "1109.3515",
    archivePrefix = "arXiv",
    primaryClass = "hep-th",
    doi = "10.1007/JHEP02(2012)126",
    journal = "JHEP",
    volume = "02",
    pages = "126",
    year = "2012"
}

@article{Hassan:2017ugh,
    author = "Hassan, S. F. and Kocic, Mikica",
    title = "{On the local structure of spacetime in ghost-free bimetric theory and massive gravity}",
    eprint = "1706.07806",
    archivePrefix = "arXiv",
    primaryClass = "hep-th",
    doi = "10.1007/JHEP05(2018)099",
    journal = "JHEP",
    volume = "05",
    pages = "099",
    year = "2018"
}

@article{Hinterbichler:2012cn,
    author = "Hinterbichler, Kurt and Rosen, Rachel A.",
    title = "{Interacting Spin-2 Fields}",
    eprint = "1203.5783",
    archivePrefix = "arXiv",
    primaryClass = "hep-th",
    doi = "10.1007/JHEP07(2012)047",
    journal = "JHEP",
    volume = "07",
    pages = "047",
    year = "2012"
}

@article{Afshar:2014dta,
    author = "Afshar, Hamid R. and Bergshoeff, Eric A. and Merbis, Wout",
    title = "{Interacting spin-2 fields in three dimensions}",
    eprint = "1410.6164",
    archivePrefix = "arXiv",
    primaryClass = "hep-th",
    reportNumber = "UG-14-18",
    doi = "10.1007/JHEP01(2015)040",
    journal = "JHEP",
    volume = "01",
    pages = "040",
    year = "2015"
}

@article{deRham:2015cha,
    author = "de Rham, Claudia and Tolley, Andrew J.",
    title = "{Vielbein to the rescue? Breaking the symmetric vielbein condition in massive gravity and multigravity}",
    eprint = "1505.01450",
    archivePrefix = "arXiv",
    primaryClass = "hep-th",
    doi = "10.1103/PhysRevD.92.024024",
    journal = "Phys. Rev. D",
    volume = "92",
    number = "2",
    pages = "024024",
    year = "2015"
}

@article{Flinckman:2025bje,
    author = "Flinckman, J. and Hassan, S. F.",
    title = "{Existence of ghost-eliminating constraints in multivielbein theory}",
    eprint = "2510.03014",
    archivePrefix = "arXiv",
    primaryClass = "hep-th",
    doi = "10.1007/JHEP07(2026)137",
    journal = "JHEP",
    volume = "07",
    pages = "137",
    year = "2026"
}

@article{Flinckman:2024zpb,
    author = "Flinckman, J. and Hassan, S. F.",
    title = "{Mass spectrum {\&} linear perturbations of ghost-free multi-spin-2 theory}",
    eprint = "2410.09439",
    archivePrefix = "arXiv",
    primaryClass = "hep-th",
    doi = "10.1007/JHEP02(2025)176",
    journal = "JHEP",
    volume = "02",
    pages = "176",
    year = "2025"
}

@article{Hassan:2018mcw,
    author = "Hassan, S. F. and Schmidt-May, Angnis",
    title = "{Interactions of multiple spin-2 fields beyond pairwise couplings}",
    eprint = "1804.09723",
    archivePrefix = "arXiv",
    primaryClass = "hep-th",
    reportNumber = "MPP-2018-66",
    doi = "10.1103/PhysRevLett.122.251101",
    journal = "Phys. Rev. Lett.",
    volume = "122",
    number = "25",
    pages = "251101",
    year = "2019"
}

@article{Flinckman:2026non,
    author = "Flinckman, Joakim and Hassan, S. F.",
    title = "{On the Uniqueness of Ghost-Free Multi-Gravity -- II: Constraining antisymmetrised multi spin-2 interactions}",
    eprint = "2604.07625",
    archivePrefix = "arXiv",
    primaryClass = "hep-th",
    month = "4",
    year = "2026"
}

@article{Khosravi:2011zi,
    author = "Khosravi, Nima and Rahmanpour, Nafiseh and Sepangi, Hamid Reza and Shahidi, Shahab",
    title = "{Multi-Metric Gravity via Massive Gravity}",
    eprint = "1111.5346",
    archivePrefix = "arXiv",
    primaryClass = "hep-th",
    doi = "10.1103/PhysRevD.85.024049",
    journal = "Phys. Rev. D",
    volume = "85",
    pages = "024049",
    year = "2012"
}

@article{Molaee:2019knc,
    author = "Molaee, Zahra and Shirzad, Ahmad",
    title = "{Hamiltonian formalism of the ghost free Tri(-Multi)gravity theory}",
    eprint = "1908.05041",
    archivePrefix = "arXiv",
    primaryClass = "hep-th",
    doi = "10.1088/1361-6382/abda01",
    journal = "Class. Quant. Grav.",
    volume = "38",
    number = "6",
    pages = "065006",
    year = "2021"
}

@article{Dokhani:2020jxb,
    author = "Dokhani, Ali and Molaee, Zahra and Shirzad, Ahmad",
    title = "{Gauge generator for bi-gravity and multi-gravity models}",
    eprint = "2001.10947",
    archivePrefix = "arXiv",
    primaryClass = "hep-th",
    doi = "10.1016/j.nuclphysb.2021.115360.",
    journal = "Nucl. Phys. B",
    volume = "966",
    pages = "115360",
    year = "2021"
}

@article{vonStrauss:2011mq,
    author = "von Strauss, Mikael and Schmidt-May, Angnis and Enander, Jonas and Mortsell, Edvard and Hassan, S. F.",
    title = "{Cosmological Solutions in Bimetric Gravity and their Observational Tests}",
    eprint = "1111.1655",
    archivePrefix = "arXiv",
    primaryClass = "gr-qc",
    doi = "10.1088/1475-7516/2012/03/042",
    journal = "JCAP",
    volume = "03",
    pages = "042",
    year = "2012"
}

@article{Akrami:2012vf,
    author = "Akrami, Yashar and Koivisto, Tomi S. and Sandstad, Marit",
    title = "{Accelerated expansion from ghost-free bigravity: a statistical analysis with improved generality}",
    eprint = "1209.0457",
    archivePrefix = "arXiv",
    primaryClass = "astro-ph.CO",
    doi = "10.1007/JHEP03(2013)099",
    journal = "JHEP",
    volume = "03",
    pages = "099",
    year = "2013"
}

@article{Akrami:2015qga,
    author = {Akrami, Yashar and Hassan, S. F. and K{\"o}nnig, Frank and Schmidt-May, Angnis and Solomon, Adam R.},
    title = "{Bimetric gravity is cosmologically viable}",
    eprint = "1503.07521",
    archivePrefix = "arXiv",
    primaryClass = "gr-qc",
    reportNumber = "NORDITA-2015-31",
    doi = "10.1016/j.physletb.2015.06.062",
    journal = "Phys. Lett. B",
    volume = "748",
    pages = "37--44",
    year = "2015"
}

@article{Volkov:2011an,
    author = "Volkov, Mikhail S.",
    title = "{Cosmological solutions with massive gravitons in the bigravity theory}",
    eprint = "1110.6153",
    archivePrefix = "arXiv",
    primaryClass = "hep-th",
    doi = "10.1007/JHEP01(2012)035",
    journal = "JHEP",
    volume = "01",
    pages = "035",
    year = "2012"
}

@article{Comelli:2011zm,
    author = "Comelli, D. and Crisostomi, M. and Nesti, F. and Pilo, L.",
    title = "{FRW Cosmology in Ghost Free Massive Gravity}",
    eprint = "1111.1983",
    archivePrefix = "arXiv",
    primaryClass = "hep-th",
    doi = "10.1007/JHEP03(2012)067",
    journal = "JHEP",
    volume = "03",
    pages = "067",
    year = "2012",
    note = "[Erratum: JHEP 06, 020 (2012)]"
}

@article{Koennig:2013fdo,
    author = "Koennig, Frank and Patil, Aashay and Amendola, Luca",
    title = "{Viable cosmological solutions in massive bimetric gravity}",
    eprint = "1312.3208",
    archivePrefix = "arXiv",
    primaryClass = "astro-ph.CO",
    doi = "10.1088/1475-7516/2014/03/029",
    journal = "JCAP",
    volume = "03",
    pages = "029",
    year = "2014"
}

@article{Hogas:2021saw,
    author = {H{\"o}g{\r{a}}s, Marcus and M{\"o}rtsell, Edvard},
    title = "{Constraints on bimetric gravity from Big Bang nucleosynthesis}",
    eprint = "2106.09030",
    archivePrefix = "arXiv",
    primaryClass = "astro-ph.CO",
    doi = "10.1088/1475-7516/2021/11/001",
    journal = "JCAP",
    volume = "11",
    pages = "001",
    year = "2021"
}

@article{Hogas:2021lns,
    author = {H{\"o}g{\r{a}}s, Marcus and M{\"o}rtsell, Edvard},
    title = "{Constraints on bimetric gravity. Part II. Observational constraints}",
    eprint = "2101.08795",
    archivePrefix = "arXiv",
    primaryClass = "gr-qc",
    doi = "10.1088/1475-7516/2021/05/002",
    journal = "JCAP",
    volume = "05",
    pages = "002",
    year = "2021"
}

@article{Hogas:2025ahb,
    author = {H{\"o}g{\r{a}}s, Marcus and M{\"o}rtsell, Edvard},
    title = "{Bimetric gravity improves the fit to DESI BAO and eases the Hubble tension}",
    eprint = "2507.03743",
    archivePrefix = "arXiv",
    primaryClass = "astro-ph.CO",
    doi = "10.1103/zz5k-kzzk",
    journal = "Phys. Rev. D",
    volume = "112",
    number = "10",
    pages = "103515",
    year = "2025"
}

@article{Hassan:2012wr,
    author = "Hassan, S. F. and Schmidt-May, Angnis and von Strauss, Mikael",
    title = "{On Consistent Theories of Massive Spin-2 Fields Coupled to Gravity}",
    eprint = "1208.1515",
    archivePrefix = "arXiv",
    primaryClass = "hep-th",
    doi = "10.1007/JHEP05(2013)086",
    journal = "JHEP",
    volume = "05",
    pages = "086",
    year = "2013"
}

@article{Wood:2025ujj,
    author = "Wood, Kieran",
    title = "{New formalism for perturbations of massive gravity theories around arbitrary background spacetimes}",
    eprint = "2509.15055",
    archivePrefix = "arXiv",
    primaryClass = "hep-th",
    doi = "10.1103/vspq-pkfy",
    journal = "Phys. Rev. D",
    volume = "113",
    number = "8",
    pages = "084054",
    year = "2026"
}

@article{Aoki:2016zgp,
    author = "Aoki, Katsuki and Mukohyama, Shinji",
    title = "{Massive gravitons as dark matter and gravitational waves}",
    eprint = "1604.06704",
    archivePrefix = "arXiv",
    primaryClass = "hep-th",
    doi = "10.1103/PhysRevD.94.024001",
    journal = "Phys. Rev. D",
    volume = "94",
    number = "2",
    pages = "024001",
    year = "2016"
}

@article{Babichev:2016hir,
    author = {Babichev, Eugeny and Marzola, Luca and Raidal, Martti and Schmidt-May, Angnis and Urban, Federico and Veerm{\"a}e, Hardi and von Strauss, Mikael},
    title = "{Bigravitational origin of dark matter}",
    eprint = "1604.08564",
    archivePrefix = "arXiv",
    primaryClass = "hep-ph",
    reportNumber = "LPT-ORSAY-16-75",
    doi = "10.1103/PhysRevD.94.084055",
    journal = "Phys. Rev. D",
    volume = "94",
    number = "8",
    pages = "084055",
    year = "2016"
}

@article{Babichev:2016bxi,
    author = {Babichev, Eugeny and Marzola, Luca and Raidal, Martti and Schmidt-May, Angnis and Urban, Federico and Veerm{\"a}e, Hardi and von Strauss, Mikael},
    title = "{Heavy spin-2 Dark Matter}",
    eprint = "1607.03497",
    archivePrefix = "arXiv",
    primaryClass = "hep-th",
    reportNumber = "LPT-ORSAY-16-60",
    doi = "10.1088/1475-7516/2016/09/016",
    journal = "JCAP",
    volume = "09",
    pages = "016",
    year = "2016"
}

@article{Luben:2018ekw,
    author = {L{\"u}ben, Marvin and M{\"o}rtsell, Edvard and Schmidt-May, Angnis},
    title = "{Bimetric cosmology is compatible with local tests of gravity}",
    eprint = "1812.08686",
    archivePrefix = "arXiv",
    primaryClass = "gr-qc",
    reportNumber = "MPP-2018-303",
    doi = "10.1088/1361-6382/ab4f9b",
    journal = "Class. Quant. Grav.",
    volume = "37",
    number = "4",
    pages = "047001",
    year = "2020"
}

@article{SupernovaCosmologyProject:1998vns,
    author = "Perlmutter, S. and others",
    collaboration = "Supernova Cosmology Project",
    title = "{Measurements of $\Omega$ and $\Lambda$ from 42 High Redshift Supernovae}",
    eprint = "astro-ph/9812133",
    archivePrefix = "arXiv",
    reportNumber = "LBNL-41801, LBL-41801",
    doi = "10.1086/307221",
    journal = "Astrophys. J.",
    volume = "517",
    pages = "565--586",
    year = "1999"
}

@article{SupernovaSearchTeam:1998fmf,
    author = "Riess, Adam G. and others",
    collaboration = "Supernova Search Team",
    title = "{Observational evidence from supernovae for an accelerating universe and a cosmological constant}",
    eprint = "astro-ph/9805201",
    archivePrefix = "arXiv",
    doi = "10.1086/300499",
    journal = "Astron. J.",
    volume = "116",
    pages = "1009--1038",
    year = "1998"
}

@article{Planck:2018vyg,
    author = "Aghanim, N. and others",
    collaboration = "Planck",
    title = "{Planck 2018 results. VI. Cosmological parameters}",
    eprint = "1807.06209",
    archivePrefix = "arXiv",
    primaryClass = "astro-ph.CO",
    doi = "10.1051/0004-6361/201833910",
    journal = "Astron. Astrophys.",
    volume = "641",
    pages = "A6",
    year = "2020",
    note = "[Erratum: Astron.Astrophys. 652, C4 (2021)]"
}

@article{Riess:2021jrx,
    author = "Riess, Adam G. and others",
    title = "{A Comprehensive Measurement of the Local Value of the Hubble Constant with 1 km s$^{-1}$ Mpc$^{-1}$ Uncertainty from the Hubble Space Telescope and the SH0ES Team}",
    eprint = "2112.04510",
    archivePrefix = "arXiv",
    primaryClass = "astro-ph.CO",
    doi = "10.3847/2041-8213/ac5c5b",
    journal = "Astrophys. J. Lett.",
    volume = "934",
    number = "1",
    pages = "L7",
    year = "2022"
}

@article{Rubin:1980zd,
    author = "Rubin, V. C. and Thonnard, N. and Ford, Jr., W. K.",
    title = "{Rotational properties of 21 SC galaxies with a large range of luminosities and radii, from NGC 4605 /R = 4kpc/ to UGC 2885 /R = 122 kpc/}",
    doi = "10.1086/158003",
    journal = "Astrophys. J.",
    volume = "238",
    pages = "471",
    year = "1980"
}

@article{Clowe:2006eq,
    author = "Clowe, Douglas and Bradac, Marusa and Gonzalez, Anthony H. and Markevitch, Maxim and Randall, Scott W. and Jones, Christine and Zaritsky, Dennis",
    title = "{A direct empirical proof of the existence of dark matter}",
    eprint = "astro-ph/0608407",
    archivePrefix = "arXiv",
    reportNumber = "SLAC-PUB-12078",
    doi = "10.1086/508162",
    journal = "Astrophys. J. Lett.",
    volume = "648",
    pages = "L109--L113",
    year = "2006"
}

@article{Bertone:2016nfn,
    author = "Bertone, Gianfranco and Hooper, Dan",
    title = "{History of dark matter}",
    eprint = "1605.04909",
    archivePrefix = "arXiv",
    primaryClass = "astro-ph.CO",
    reportNumber = "FERMILAB-PUB-16-157-A",
    doi = "10.1103/RevModPhys.90.045002",
    journal = "Rev. Mod. Phys.",
    volume = "90",
    number = "4",
    pages = "045002",
    year = "2018"
}

@article{DeFelice:2015hla,
    author = "De Felice, Antonio and Mukohyama, Shinji",
    title = "{Minimal theory of massive gravity}",
    eprint = "1506.01594",
    archivePrefix = "arXiv",
    primaryClass = "hep-th",
    reportNumber = "YITP-15-48, IPMU15-0081",
    doi = "10.1016/j.physletb.2015.11.050",
    journal = "Phys. Lett. B",
    volume = "752",
    pages = "302--305",
    year = "2016"
}

@article{DeFelice:2020ecp,
    author = "De Felice, Antonio and Larrouturou, Fran{\c{c}}ois and Mukohyama, Shinji and Oliosi, Michele",
    title = "{Minimal Theory of Bigravity: construction and cosmology}",
    eprint = "2012.01073",
    archivePrefix = "arXiv",
    primaryClass = "gr-qc",
    reportNumber = "YITP-20-157, IPMU20-0126",
    doi = "10.1088/1475-7516/2021/04/015",
    journal = "JCAP",
    volume = "04",
    pages = "015",
    year = "2021"
}

@article{Cusin:2015tmf,
    author = "Cusin, Giulia and Durrer, Ruth and Guarato, Pietro and Motta, Mariele",
    title = "{A general mass term for bigravity}",
    eprint = "1512.02131",
    archivePrefix = "arXiv",
    primaryClass = "astro-ph.CO",
    doi = "10.1088/1475-7516/2016/04/051",
    journal = "JCAP",
    volume = "04",
    pages = "051",
    year = "2016"
}

@article{DeFelice:2012mx,
    author = "De Felice, Antonio and Gumrukcuoglu, A. Emir and Mukohyama, Shinji",
    title = "{Massive gravity: nonlinear instability of the homogeneous and isotropic universe}",
    eprint = "1206.2080",
    archivePrefix = "arXiv",
    primaryClass = "hep-th",
    reportNumber = "IPMU12-0118",
    doi = "10.1103/PhysRevLett.109.171101",
    journal = "Phys. Rev. Lett.",
    volume = "109",
    pages = "171101",
    year = "2012"
}

@article{Max:2017flc,
    author = "Max, Kevin and Platscher, Moritz and Smirnov, Juri",
    title = "{Gravitational Wave Oscillations in Bigravity}",
    eprint = "1703.07785",
    archivePrefix = "arXiv",
    primaryClass = "gr-qc",
    doi = "10.1103/PhysRevLett.119.111101",
    journal = "Phys. Rev. Lett.",
    volume = "119",
    number = "11",
    pages = "111101",
    year = "2017"
}

@article{Max:2017kdc,
    author = "Max, Kevin and Platscher, Moritz and Smirnov, Juri",
    title = "{Decoherence of Gravitational Wave Oscillations in Bigravity}",
    eprint = "1712.06601",
    archivePrefix = "arXiv",
    primaryClass = "gr-qc",
    doi = "10.1103/PhysRevD.97.064009",
    journal = "Phys. Rev. D",
    volume = "97",
    number = "6",
    pages = "064009",
    year = "2018"
}

@article{Cembranos:2026cqk,
    author = "Cembranos, Jose A. R. and Cendal, {\'A}lvaro and Villarrubia-Rojo, Hector",
    title = "{Effects of massive spin-2 fields on gravitational wave propagation}",
    eprint = "2601.15201",
    archivePrefix = "arXiv",
    primaryClass = "gr-qc",
    reportNumber = "IPARCOS-UCM-26-002",
    doi = "10.1088/1475-7516/2026/06/008",
    journal = "JCAP",
    volume = "06",
    pages = "008",
    year = "2026"
}

@article{Vainshtein:1972sx,
    author = "Vainshtein, A. I.",
    title = "{To the problem of nonvanishing gravitation mass}",
    doi = "10.1016/0370-2693(72)90147-5",
    journal = "Phys. Lett. B",
    volume = "39",
    pages = "393--394",
    year = "1972"
}

@article{Babichev:2013pfa,
    author = "Babichev, Eugeny and Crisostomi, Marco",
    title = "{Restoring general relativity in massive bigravity theory}",
    eprint = "1307.3640",
    archivePrefix = "arXiv",
    primaryClass = "gr-qc",
    reportNumber = "LPT-ORSAY-13-72",
    doi = "10.1103/PhysRevD.88.084002",
    journal = "Phys. Rev. D",
    volume = "88",
    number = "8",
    pages = "084002",
    year = "2013"
}

@article{Mortsell:2015exa,
    author = "Mortsell, E. and Enander, J.",
    title = "{Scalar instabilities in bimetric gravity: The Vainshtein mechanism and structure formation}",
    eprint = "1506.04977",
    archivePrefix = "arXiv",
    primaryClass = "astro-ph.CO",
    doi = "10.1088/1475-7516/2015/10/044",
    journal = "JCAP",
    volume = "10",
    pages = "044",
    year = "2015"
}

@article{Luben:2019yyx,
    author = {L{\"u}ben, Marvin and Schmidt-May, Angnis and Smirnov, Juri},
    title = "{Vainshtein Screening in Bimetric Cosmology}",
    eprint = "1912.09449",
    archivePrefix = "arXiv",
    primaryClass = "gr-qc",
    doi = "10.1103/PhysRevD.102.123529",
    journal = "Phys. Rev. D",
    volume = "102",
    pages = "123529",
    year = "2020"
}

\end{document}